
\input lanlmac
\input epsf
\input amssym
\noblackbox
\figno=0
\def\fig#1#2#3{
\par\begingroup\parindent=0pt\leftskip=1cm\rightskip=1cm\parindent=0pt
\baselineskip=11pt
\global\advance\figno by 1
\midinsert
\epsfxsize=#3
\centerline{\epsfbox{#2}}
\vskip 12pt
{\bf Fig. \the\figno:} #1\par
\endinsert\endgroup\par
}
\def\figlabel#1{\xdef#1{\the\figno}}
\def\encadremath#1{\vbox{\hrule\hbox{\vrule\kern8pt\vbox{\kern8pt
\hbox{$\displaystyle #1$}\kern8pt}
\kern8pt\vrule}\hrule}}

\def\figtwo#1#2#3#4#5#6{
\par\begingroup\parindent=0pt\leftskip=1cm\rightskip=1cm\parindent=0pt
\baselineskip=11pt
\global\advance\figno by 1
\midinsert
\centerline{\epsfxsize=#3 a)\epsfbox{#2}\hskip #4 \epsfxsize=#6 b)\epsfbox{#5}}
\vskip 12pt
{\bf Fig. \the\figno:} #1\par
\endinsert\endgroup\par
}
\def\figlabel#1{\xdef#1{\the\figno}}
\def\encadremath#1{\vbox{\hrule\hbox{\vrule\kern8pt\vbox{\kern8pt
\hbox{$\displaystyle #1$}\kern8pt}
\kern8pt\vrule}\hrule}}

\def\hf{ \frac{1}{2}}
\def\half{ {1\over 2} }
\def\encadremath#1{\vbox{\hrule\hbox{\vrule\kern8pt\vbox{\kern8pt
 \hbox{$\displaystyle #1$}\kern8pt}
 \kern8pt\vrule}\hrule}}


\def\rlx{\relax\leavevmode}
\def\inbar{\vrule height1.5ex width.4pt depth0pt}
\def\IC{\relax\,\hbox{$\inbar\kern-.3em{\rm C}$}}
\def\IN{\relax{\rm I\kern-.18em N}}
\def\IP{\relax{\rm I\kern-.18em P}}

\def\ZZ{\rlx\leavevmode\ifmmode\mathchoice{\hbox{\cmss Z\kern-.4em Z}}
{\hbox{\cmss Z\kern-.4em Z}}{\lower.9pt\hbox{\cmsss Z\kern-.36em Z}}
{\lower1.2pt\hbox{\cmsss Z\kern-.36em Z}}\else{\cmss Z\kern-.4em Z}\fi}
\def\IZ{\relax\ifmmode\mathchoice
{\hbox{\cmss Z\kern-.4em Z}}{\hbox{\cmss Z\kern-.4em Z}}
{\lower.9pt\hbox{\cmsss Z\kern-.4em Z}}
{\lower1.2pt\hbox{\cmsss Z\kern-.4em Z}}\else{\cmss Z\kern-.4em Z}\fi}
\def\IB{\relax{\rm I\kern-.18em B}}
\def\IC{{\relax\hbox{$\inbar\kern-.3em{\rm C}$}}}
\def\Ic{{\relax\hbox{$\inbar\kern-.22em{\rm c}$}}}
\def\ID{\relax{\rm I\kern-.18em D}}
\def\IE{\relax{\rm I\kern-.18em E}}
\def\IF{\relax{\rm I\kern-.18em F}}
\def\IG{\relax\hbox{$\inbar\kern-.3em{\rm G}$}}
\def\IGa{\relax\hbox{${\rm I}\kern-.18em\Gamma$}}
\def\IH{\relax{\rm I\kern-.18em H}}
\def\II{\relax{\rm I\kern-.18em I}}
\def\IK{\relax{\rm I\kern-.18em K}}
\def\IP{\relax{\rm I\kern-.18em P}}
\def\IR{\relax{\rm I\kern-.18em R}}

\font\cmss=cmss10
\font\cmsss=cmss10 at 7pt


\def\eqnn#1{\xdef
#1{(\secsym\the\meqno)}\writedef{#1\leftbracket#1}%
\global\advance\meqno by1\wrlabeL#1}
\def\eqna#1{\xdef
#1##1{\hbox{$(\secsym\the\meqno##1)$}}

\writedef{#1\numbersign1\leftbracket#1{\numbersign1}}%
 \global\advance\meqno by1\wrlabeL{#1$\{\}$}}
\def\eqn#1#2{\xdef
 #1{(\secsym\the\meqno)}\writedef{#1\leftbracket#1}%
 \global\advance\meqno by1$$#2\eqno#1\eqlabeL#1$$}

   \def\R{\relax{\rm I\kern-.18em R}}
   \font\cmss=cmss10 \font\cmsss=cmss10 at 7pt
   \def\Z{\relax\ifmmode\mathchoice
   {\hbox{\cmss Z\kern-.4em Z}}{\hbox{\cmss Z\kern-.4em Z}}
   {\lower.9pt\hbox{\cmsss Z\kern-.4em Z}}
   {\lower1.2pt\hbox{\cmsss Z\kern-.4em Z}}\else{\cmss Z\kern-.4em
   Z}\fi}
 \def\bigone{\hbox{1\kern -.23em {\rm l}}}


 \chardef\tempcat=\the\catcode`\@ \catcode`\@=11
 \def\cyracc{\def\u##1{\if \i##1\accent"24 i%
     \else \accent"24 ##1\fi }}
 \newfam\cyrfam
 \font\tencyr=wncyr10
 \def\cyr{\fam\cyrfam\tencyr\cyracc}

\def\frac#1#2{{\textstyle{#1\over#2}}}



\def\Tr{\,{\rm Tr}\, }
\def\det{\,{\rm det}\, }

\def\Im{\,{\rm Im}\, }

\def\({\left(}
\def\){\right)}
\def\[{\left[}
\def\]{\right]}
\def\p{\partial}

\def\11{1\!\! 1}
\def\tint{{\int\!\!\!\int}}


\def\inbar{\,\vrule height1.5ex width.4pt depth0pt}
\def\IB{\relax{\rm I\kern-.18em B}}
\def\IC{\relax\hbox{$\inbar\kern-.3em{\rm C}$}}
\def\ID{\relax{\rm I\kern-.18em D}}
\def\IE{\relax{\rm I\kern-.18em E}}
\def\IF{\relax{\rm I\kern-.18em F}}
\def\IG{\relax\hbox{$\inbar\kern-.3em{\rm G}$}}
\def\IH{\relax{\rm I\kern-.18em H}}
\def\II{\relax{\rm I\kern-.18em I}}
\def\IK{\relax{\rm I\kern-.18em K}}
\def\IL{\relax{\rm I\kern-.18em L}}
\def\IM{\relax{\rm I\kern-.18em M}}
\def\IN{\relax{\rm I\kern-.18em N}}
\def\IO{\relax\hbox{$\inbar\kern-.3em{\rm O}$}}
\def\IP{\relax{\rm I\kern-.18em P}}
\def\IQ{\relax\hbox{$\inbar\kern-.3em{\rm Q}$}}
\def\IR{\relax{\rm I\kern-.18em R}}
\font\cmss=cmss10 \font\cmsss=cmss10 at 7pt
\def\IZ{\relax\ifmmode\mathchoice
{\hbox{\cmss Z\kern-.4em Z}}{\hbox{\cmss Z\kern-.4em Z}}
{\lower.9pt\hbox{\cmsss Z\kern-.4em Z}}
{\lower1.2pt\hbox{\cmsss Z\kern-.4em Z}}\else{\cmss Z\kern-.4em Z}\fi}
\def\IGa{\relax\hbox{${\rm I}\kern-.18em\Gamma$}}
\def\IPi{\relax\hbox{${\rm I}\kern-.18em\Pi$}}

 \def\({\left(}
 \def\){\right)}
 \def\<{\left\langle}
 \def\>{\right\rangle}


\def\b{\beta}
\def\g{\gamma}
\def\d{\delta}
\def\e{\epsilon}
\def\eps{\varepsilon}

\def\th{\theta}

\def\l{\lambda}

\def\vr{\varrho}

\def\t{\tau}
\def\u{\upsilon }

\def\vp{\varphi}

\def\G{\Gamma}

\def\L{\Lambda}

\def\o{\omega }


   \def\CF {{\cal F}}

   \def\CM {{\cal M}}
   
   \def\CO {{\cal O}}

   \def\CS {{\cal S}}
   \def\CT {{\cal T}}
   \def\CU {{\cal U}}

   \def\CZ {{\cal Z}}

\def\Xb{{\bf X}}

\def\Pb{{\bf P}}
\def\Ab{{\bf A}}





\def\Xp{X_{_{+}}}
\def\Xm{X_{_{-}}}
\def\Xpm{X_{_{\pm}}}

\def\Xbp{{\bf X}_{_{+}}}
\def\Xbm{{\bf X}_{_{-}}}
\def\Xbpm{{\bf X}_{_{\pm}}}

\def\xp{x_{_{+}}}
\def\xm{x_{_{-}}}
\def\xpm{x_{_{\pm}}}
\def\xmp{x_{_{\mp}}}

\def\yp{y_{_{+}}}
\def\ym{y_{_{-}}}
\def\ypm{y_{_{\pm}}}

\def\txp{X_{_{+}}}
\def\txm{X_{_{-}}}
\def\txpm{X_{_{\pm}}}
\def\txmp{X_{_{\mp}}}

\def\pxp{x_{_{+}}^{_{'}}}
\def\pxm{x_{_{-}}^{_{'}}}
\def\pxpm{x_{_{\pm}}^{_{'}}}


\def\Pe{\Psi^{_{E}}}

\def\pse{ \psi^{_{E}} }
\def\pee{\psi^{_{E'}}}

\def\psip{\psi_{_{+}}}
\def\psim{\psi_{_{-}}}

\def\Psp{\Psi_{_{+}}}
\def\Psm{\Psi_{_{-}}}
\def\Pspm{\Psi_{_{\pm}}}

\def\Php{\Phi_{_{+}}}
\def\Phm{\Phi_{_{-}}}
\def\Phpm{\Phi_{_{\pm}}}

\def\Pep{\Pe_{_{+}}}
\def\Pem{\Pe_{_{-}}}
\def\Pepm{\Pe_{_{\pm}}}

\def\psep{\pse_{_{+}}}
\def\psem{\pse_{_{-}}}
\def\psepm{\pse_{_{\pm }}}

\def\pseep{\pee_{_{+}}}

\def\Dp{D_{_{+}}}
\def\Dm{D_{_{-}}}
\def\Dpm{D_{_{\pm}}}


\def\XX{X}
\def\oR{\frac{1}{R}}

\def\under#1#2{\mathop{#1}\limits_{#2}}

\def\FSL{\CF}

\def\gst{g_{\rm str}}

\def\ctf{\sqrt{\Lambda}}

\def\Pfzz{\Phi_{_{FZZT}}}

\def\gf{\gamma_{\rm F}}

\def\cqg#1#2#3{{\it Class. Quantum Grav.} {\bf #1} (#2) #3}
\def\np#1#2#3{{\it Nucl. Phys.} {\bf B#1} (#2) #3}
\def\pl#1#2#3{{\it Phys. Lett. }{\bf B#1} (#2) #3}

\def\physrev#1#2#3{{\it Phys. Rev.} {\bf D#1} (#2) #3}

\def\physrep#1#2#3{{\it Phys. Rep.} {\bf #1} (#2) #3}

\def\jhep#1#2#3{{\it JHEP} {\bf #1} (#2) #3}
\def\hepth#1{[arXiv:hep-th/#1]}

 \lref\MOORE{
G. Moore, ``Gravitational phase transitions and the sine-Gordon model", 
\hepth{9203061}.}
  
\lref\DMP{
R. Dijkgraaf, G. Moore, and M.R. Plesser,
``The partition function of 2d string theory'',
\np{394}{1993}{356}, \hepth{9208031}.}

\lref\UT{
K. Ueno and K. Takasaki, ``Toda Lattice Hierarchy":
in `Group representations and systems of differential equations',
{\it Adv. Stud. Pure Math.} {\bf 4} (1984) 1.}

\lref\Takasak{
K. Takasaki,
{\it Adv. Stud. Pure Math.} {\bf 4} (1984) 139.}

\lref\TeodorescuYA{
R.~Teodorescu, E.~Bettelheim, O.~Agam, A.~Zabrodin, and P.~Wiegmann,
``Semiclassical evolution of the spectral curve in the normal random matrix
ensemble as Whitham hierarchy,''
\np{700}{2004}{521}, \hepth{0407017}].
}

\lref\WiegmannXF{
P.~Wiegmann and A.~Zabrodin,
``Large N expansion for normal and complex matrix ensembles,''
[arXiv:hep-th/0309253].
}

\lref\MooreSF{
G.W.~Moore, ``Double scaled field theory at c = 1,''
\np{368}{1992}{557}.
}

\lref\Hos{
K.~Hosomichi,
``Bulk-Boundary Propagator in Liouville Theory on a Disk,''
\jhep{0111}{2001}{044},
[arXiv:hep-th/0108093].}

\lref\TeschPons{
B.~Ponsot and J.~Teschner,
``Boundary Liouville field theory: Boundary three point function,''
\np{622}{2002}{309}, [arXiv:hep-th/0110244].}

\lref\Pons{
B.~Ponsot, ``Liouville Theory on the Pseudosphere: Bulk-Boundary Structure
Constant,''
\pl{588}{2004}{105}, [arXiv:hep-th/0309211].}

\lref\McGreevyKB{
J.~McGreevy and H.~Verlinde,
``Strings from tachyons: The $c = 1$ matrix reloaded,''
\jhep{0312}{2003}{054}, [arXiv:hep-th/0304224].
}

\lref\JevickiQN{
A.~Jevicki,
``Development in 2-d string theory,''
[arXiv:hep-th/9309115].
}

\lref\MooreIR{
G.~W.~Moore, N.~Seiberg, and M.~Staudacher,
``From loops to states in 2-D quantum gravity,''
\np{362}{1991}{665}.
}

\lref\KlebanovKM{
I.R.~Klebanov, J.~Maldacena, and N.~Seiberg,
``D-brane decay in two-dimensional string theory,''
\jhep{0307}{2003}{045}, [arXiv:hep-th/0305159].
}

\lref\McGreevyEP{
J.~McGreevy, J.~Teschner, and H.~Verlinde,
``Classical and quantum D-branes in 2D string theory,''
\jhep{0401}{2004}{039}, [arXiv:hep-th/0305194].
}

\lref\Barnes{
E.W. Barnes, ``The theory of the double gamma function",
{\it Philos. Trans. Roy. Soc.} A {\bf 196} (1901) 265.}

\lref\KAK{
S.Yu. Alexandrov, V.A. Kazakov, and D. Kutasov,
``Non-Perturbative Effects in Matrix Models and D-branes,''
\jhep{0309}{2003}{057}, [arXiv:hep-th/0306177].
}

\lref\SeibergNM{
N.~Seiberg and D.~Shih,
``Branes, rings and matrix models in minimal (super)string theory,''
\jhep{0402}{2004}{021}, [arXiv:hep-th/0312170].
}

\lref\KutasovFG{
D.~Kutasov, K.~Okuyama, J.~Park, N.~Seiberg, and D.~Shih,
``Annulus amplitudes and ZZ branes in minimal string theory,''
\jhep{0408}{2004}{026}, [arXiv:hep-th/0406030].
}

\lref\CallanPT{
C.G.~.~Callan and S.R.~Coleman,
``The Fate Of The False Vacuum. 2. First Quantum Corrections,''
\physrev{16}{1977}{1762}.
}

\lref\SAmn{
S. Alexandrov,
``$(m,n)$ ZZ branes and the $c=1$ matrix model,''
\pl{604}{2004}{115}, [arXiv:hep-th/0310135].
}

\lref\KazakovDU{
V.A.~Kazakov and I.K.~Kostov,
``Instantons in non-critical strings from the two-matrix model,''
[arXiv:hep-th/0403152].
}

\lref\SAcurve{
S. Alexandrov,
``D-branes and complex curves in $c=1$ string theory,''
\jhep{0405}{2004}{025}, [arXiv:hep-th/0403116].
}

\lref\HanadaIM{
M.~Hanada, M.~Hayakawa, N.~Ishibashi, H.~Kawai, T.~Kuroki, Y.~Matsuo, 
and T.~Tada,
``Loops versus matrices: The nonperturbative aspects of noncritical
string,''
{\it Prog. Theor. Phys.} {\bf 112} (2004) 131, [arXiv:hep-th/0405076].
}

\lref\KostovTK{
I.K.~Kostov, ``Integrable flows in c = 1 string theory,''
{\it J.\ Phys.}\ A {\bf 36} (2003) 3153 
(Annales Henri Poincare {\bf 4} (2003) S825), [arXiv:hep-th/0208034].
}

\lref\KostovWV{
I.K.~Kostov,
``String equation for string theory on a circle,''
\np{624}{2002}{146}, [arXiv:hep-th/0107247].
}

\lref\AKK{
S.Yu. Alexandrov, V.A. Kazakov, and I.K. Kostov,
``Time-dependent backgrounds of 2D string theory,''
\np{640}{2002}{119}, [arXiv:hep-th/0205079].
}

\lref\Witgr{
E. Witten, ``Ground Ring of two dimensional string theory,''
\np{373}{1992}{187}, [arXiv:hep-th/9108004].}

\lref\KostovGR{
I.K.~Kostov,
``Boundary ground ring in 2D string theory,''
\np{689}{2004}{3}, [arXiv:hep-th/0312301].}

\lref\SAthese{
S. Alexandrov,
``Matrix quantum mechanics and two-dimensional string theory
in non-trivial backgrounds,'' PhD thesis,
[arXiv:hep-th/0311273].
}

\lref\AK{
S. Alexandrov and V. Kazakov,
``Correlators in 2D string theory with vortex condensation,''
\np{610}{2001}{77}, [arXiv:hep-th/0104094].
}

\lref\AKKNMM{
S.Yu. Alexandrov, V.A. Kazakov, and I.K. Kostov,
``2D String Theory as Normal Matrix Model,''
\np{667}{2003}{90}, [arXiv:hep-th/0302106].
}

\lref\HSU{
E. Hsu and D. Kutasov, ``The Gravitational Sine-Gordon Model,''
\np{396}{1993}{693}, [arXiv:hep-th/9212023].}

\lref\deBoerHD{
J.~de Boer, A.~Sinkovics, E.~Verlinde, and J.T.~Yee,
``String interactions in c = 1 matrix model,''
\jhep{0403}{2004}{023}, [arXiv:hep-th/0312135].
}

\lref\DijkgraafVP{
R.~Dijkgraaf, A.~Sinkovics, and M.~Temurhan,
``Universal correlators from geometry,''
[arXiv:hep-th/0406247].
}

\lref\BoyarskyJB{
A.~Boyarsky, B.~Kulik, and O.~Ruchayskiy,
``Classical and quantum branes in c = 1 string theory and quantum Hall
effect,''
[arXiv:hep-th/0312242].
}

\lref\GinspargIS{
P.H.~Ginsparg and G.~W.~Moore,
``Lectures On 2-D Gravity And 2-D String Theory,''
[arXiv:hep-th/9304011].
}

\lref\DiFrancescoNW{
P.~Di Francesco, P.H.~Ginsparg, and J.~Zinn-Justin,
``2-D Gravity and random matrices,''
\physrep{254}{1995}{1}, [arXiv:hep-th/9306153].
}

\lref\KlebanovQA{
I.R.~Klebanov,
``String theory in two-dimensions,''
[arXiv:hep-th/9108019].
}

\lref\ShenkerUF{
S.H.~Shenker,
``The Strength Of Nonperturbative Effects In String Theory,'' RU-90-47
{\it Presented at the Cargese Workshop on Random Surfaces,
Quantum Gravity and Strings, Cargese, France, May 28 - Jun 1, 1990}
}

\lref\DavidSK{
F.~David,
``Phases Of The Large N Matrix Model And Non-perturbative Effects In 2-D
Gravity,''
\np{348}{1991}{507}.
}

\lref\EynardSG{
B.~Eynard and J.~Zinn-Justin,
``Large order behavior of 2-D gravity coupled to d < 1 matter,''
\pl{302}{1993}{396}, [arXiv:hep-th/9301004].
}

\lref\MartinecKA{
E.J.~Martinec,
``The annular report on non-critical string theory,''
[arXiv:hep-th/0305148].
}

\lref\DouglasUP{
M.R.~Douglas, I.R.~Klebanov, D.~Kutasov, J.~Maldacena, E.~Martinec, and
N.~Seiberg,
``A new hat for the c = 1 matrix model,''
[arXiv:hep-th/0307195].
}

\lref\KostovXI{
I.K.~Kostov,
``Conformal field theory techniques in random matrix models,''
[arXiv:hep-th/9907060].
}

\lref\DornXN{
H.~Dorn and H.J.~Otto,
``Two and three point functions in Liouville theory,''
\np{429}{1994}{375}, [arXiv:hep-th/9403141].
}

\lref\ZamolodchikovAA{
A.B.~Zamolodchikov and A.B.~Zamolodchikov,
``Structure constants and conformal bootstrap in Liouville field theory,''
\np{477}{1996}{577}, [arXiv:hep-th/9506136].
}

\lref\MartinecQT{
E.~Martinec and K.~Okuyama,
``Scattered Results in 2D String Theory,''
\jhep{0410}{2004}{065}, [arXiv:hep-th/0407136].
}

\lref\FateevIK{
V.~Fateev, A.B.~Zamolodchikov, and A.B.~Zamolodchikov,
``Boundary Liouville field theory. I: Boundary state and boundary two-point
function,''
[arXiv:hep-th/0001012].
}

\lref\TeschnerMD{
J.~Teschner,
``Remarks on Liouville theory with boundary,''
[arXiv:hep-th/0009138].
}

\lref\TeschnerRim{
J.~Teschner,
``On boundary perturbations in Liouville theory and brane dynamics in
noncritical string theories,''
\jhep{0404}{2004}{023}, [arXiv:hep-th/0308140].
}

\lref\ZamolodchikovAH{
A.B.~Zamolodchikov and A.B.~Zamolodchikov,
``Liouville field theory on a pseudosphere,''
[arXiv:hep-th/0101152].
}

\lref\TeschnerRV{
J.~Teschner,
``Liouville theory revisited,''
\cqg{18}{2001}{R153}, [arXiv:hep-th/0104158].
}

\lref\PolchinskiMB{
J.~Polchinski,
``What is string theory?,''
[arXiv:hep-th/9411028].
}

\lref\TakayanagiSM{
T.~Takayanagi and N.~Toumbas,
``A matrix model dual of type 0B string theory in two dimensions,''
\jhep{0307}{2003}{064}, [arXiv:hep-th/0307083].
}

\lref\KKK{
V.~Kazakov, I.~K.~Kostov, and D.~Kutasov,
``A matrix model for the two-dimensional black hole,''
\np{622}{2002}{141}, [arXiv:hep-th/0101011].
}

\lref\MaldacenaSN{
J.~Maldacena, G.~Moore, N.~Seiberg, and D.~Shih,
``Exact vs. Semiclassical Target Space of the Minimal String,''
[arXiv:hep-th/0408039].
}

\lref\GKVdilaton{
D. Grumiller, W. Kummer, and D.V. Vassilevich,
``Dilaton Gravity in Two Dimensions,"
\physrep{369}{2002}{327}, [arXiv:hep-th/0204253].
}

\lref\BoyarskySM{
A.~Boyarsky, V.~V.~Cheianov, and O.~Ruchayskiy,
``Fermions in the harmonic potential and string theory,''
[arXiv:hep-th/0409129].
}

\lref\PolchinskiFQ{
J.~Polchinski,
``Combinatorics Of Boundaries In String Theory,''
\physrev{50}{1994}{6041}, [arXiv:hep-th/9407031].
}

\lref\SenIV{
A.~Sen, ``Open-closed duality: Lessons from matrix model,''
{\it Mod.\ Phys.\ Lett.} A {\bf 19} (2004) 841, [arXiv:hep-th/0308068].
}

\lref\GaiottoYB{
D.~Gaiotto and L.~Rastelli,
``A paradigm of open/closed duality: Liouville D-branes and the Kontsevich
model,''
[arXiv:hep-th/0312196].
}

\lref\MartinecTD{
E.~J.~Martinec,
``Matrix models and 2D string theory,''
[arXiv:hep-th/0410136].
}

 \overfullrule=0pt

\Title{\vbox{\baselineskip12pt\hbox{SPhT-t04/137}
\hbox{ITP-UU-04/50}\hbox{SPIN-04/32}}}
{\vbox{\centerline{ Time-dependent backgrounds of 2D string theory:}  
\centerline{Non-perturbative effects}}}

\vskip -0.7cm
\centerline{Sergei Yu. Alexandrov\footnote{$^\flat$}{\tt S.Alexandrov@phys.uu.nl}}
\centerline{\it  Institute for Theoretical Physics \& Spinoza Institute, Utrecht University,}
\centerline{\it Postbus 80.195, 3508 TD Utrecht, The Netherlands}

\vskip 0.4cm

\centerline{Ivan K. Kostov\footnote{$^\ast$}{{\tt
kostov@spht.saclay.cea.fr}}\footnote{$^{\dag}$}{Associate member of
{\cyr  IYaIYaE -- BAN}, Sofia, Bulgaria }}
\centerline{\it  Service de Physique Th{\'e}orique, CNRS -- URA 2306, }
\centerline{\it C.E.A. - Saclay, F-91191 Gif-Sur-Yvette,
France
}


\vskip 1.5cm

\baselineskip=11pt
{\ninepoint
{We study the non-perturbative corrections (NPC) to the partition function 
of a compactified 2D string theory in a time-dependent background generated 
by a tachyon source. The sine-Liouville deformation of the theory  
is a particular case of such a background. We calculate the leading as well 
as the subleading NPC using the dual description of the string theory 
as matrix quantum mechanics. As in the minimal string theories, 
the NPC are classified by the double points of a complex curve.  
We calculate them by two different methods: by solving Toda equation  
and by evaluating the quasiclassical fermion wave functions. 
We show that the result can be expressed in terms of correlation functions 
of the bosonic field associated with the tachyon source
and identify the leading and the subleading corrections as the contributions 
from the one-point (disk) and two-point (annulus) correlation functions.}}

\Date{  }
\vfill
\eject

\baselineskip=14pt plus 2pt minus 2pt


\newsec{Introduction}

\noindent
The $c\le 1$ string theories, which can also be viewed as solvable
models of 2D quantum gravity, have been solved perturbatively through
their description as large $N$ matrix models (see the reviews
\refs{\DiFrancescoNW\GinspargIS-\KlebanovQA}). Originally the matrix
models were considered just as `engins' that generate planar graphs,
but very soon it became clear that the matrix models are in principle
able to give a qualitative picture of the non-perturbative
effects\foot{The first systematic study of the non-perturbative
effects in the simplest one-matrix model has been done by F. David
\DavidSK.} in these low-dimensional string theories \refs{\ShenkerUF}.
 
Recent works \refs{\McGreevyKB\MartinecKA\KlebanovKM\McGreevyEP
\KAK\SAmn\SeibergNM\SAcurve\KazakovDU\HanadaIM\KutasovFG\SenIV
\deBoerHD\GaiottoYB\MartinecTD-\MaldacenaSN} suggested
to think of the matrix models as dual open string theories in presence
of D-branes and led to an interpretation of the non-perturbative
phenomena in terms of open string world sheets with appropriate
boundary conditions. This interpretation was prepared, from the
string theory side, by the remarkable works on Liouville CFT with
boundary, which has been solved using conformal bootstrap methods
\refs{\FateevIK\TeschnerMD\ZamolodchikovAH\Hos-\TeschPons}. All
comparisons between the matrix and CFT approaches showed agreement and
led further to a new proposal for the matrix description of string
theories with world sheet supersymmetry, type 0A and 0B theories
\refs{\TakayanagiSM\DouglasUP-\MartinecQT}.

In particular, the non-perturbative corrections (NPC) to the string
partition function are believed to be produced by D-instantons
\PolchinskiFQ. On the string theory side, they were shown to be
described in terms of a boundary CFT with Dirichlet boundary
conditions on the Liouville field, known also as ZZ branes
\ZamolodchikovAH. Namely, the leading corrections are given by the
exponents of the disk partition functions with ZZ boundary conditions
\refs{\EynardSG,\KAK,\SAmn}.

For $c<1$ minimal string theories, the leading non-perturbative
effects were given a nice geometric interpretation in terms of a
complex curve describing the closed string theory background
\refs{\SeibergNM,\KazakovDU}. Each point on this Riemann surface is
associated with a FZZT brane, which implies Neumann boundary conditions
for the Liouville field \FateevIK. The ZZ branes are associated with
the double points of the Riemann surface, which can be also thought of
as vanishing $A$-cycles.  The disk partition functions with ZZ
boundary conditions are given by line integrals along the
corresponding $B$-cycles.

The generalization of these results to $c=1$ is not obvious because of
the singular character of the limit $c\to 1$ in the space of minimal
string theories. In particular, the double points
of the complex curve degenerate just to two singularities where all 
ZZ branes are situated. Therefore, in order to study the non-perturbative 
effects in $c=1$ string theories, it is helpful to perturb the theory with 
a time-dependent tachyon potential, to the effect that the degeneracy of the singular  
points at $c=1$ is lifted. 

The integrable tachyon perturbations are those with equidistant spectrum  
and it is natural to consider them in the context of a string theory with 
finite temperature. The world sheet description of such a theory is given by 
the euclidean $c=1$ string theory, defined by the world sheet action
\eqn\coneper{\eqalign{ 
S_{c=1}={1\over 4\pi}\int d^2\sigma\left[
(\partial \chi )^2 + (\partial\phi)^2 +2\hat R\phi+
\mu \phi\, e^{2\phi}+{\rm ghosts}\right]},
} 
in which the matter field $\chi$, or the local euclidean time coordinate of the string,   
is compactified: 
\eqn\compct{
\chi+2\pi R\equiv \chi.
}
We are interested in the possible deformations of this theory by 
on-mass-shell  tachyon operators\foot{In the
compactified theory, in addition to this discrete spectrum of tachyon
modes, there is a discrete set of the winding modes, or
Kosterlitz--Thouless vortices. In the following we will consider only
perturbations by tachyons; the results for perturbations by winding
modes can be obtained by T-duality.}
\eqn\vvintegr{
\CT_{p} \sim \int \! d^2 \sigma \,  e^{i  p \chi }e^{(2- |p|)\phi} 
}
with the allowed by the compactification \compct\ values of the momentum
\eqn\pnn{
p_k =k/R, \qquad k=\pm1, \pm2, \dots \, .
}
A general such deformation is achieved by adding
to the action \coneper\ a term
\eqn\PERTS{ \delta S = \sum_{k\ne 0} t_k \CT_{k/R} .
}
We will be mainly interested in sine-Liouville deformation,
which represents a perturbation by the two lowest vertex operators
$n=\pm 1$. It contains, besides the Liouville term $\mu e^{2\phi}$,
a sine-Liouville interaction 
\eqn\sinelio{ 
\delta S_{_{SL}}= \l \int \! d^2 \sigma \, \cos(\chi /R) \, 
e^{(2-{1\over R})\phi},}
where we denoted  $\l = t_1 =t_{-1}$. 

The string theory deformed by the term \PERTS\ has been studied 
using the `holographic' description provided by the Matrix Quantum 
Mechanics (MQM). The  first exact result, the expression for 
the partition function for sine-Liouville deformation, was suggested  
by G. Moore in \MOORE. Later  this result was proved and generalised  
in \refs{\KKK\AK-\AKK} using the fact that the perturbations \PERTS\ 
behave as Toda integrable flows.
The classical background for given deformation is described by a complex curve   
whose exact form depends on the coupling constants in \PERTS.
This curve is not the one describing the FZZT brane, but is obtained from 
the latter by a projection that identifies an infinite number of sheets. 
The existence of two distinct complex curves is 
a peculiarity of the $c=1$ string theory and is related with the 
logarithmic singularity of the resolvent at infinity. 
In the $c<1$ string theories there is only one such curve.
  
The leading non-perturbative effects in presence of sine-Liouville  
deformation were studied in \refs{\KAK,\SAmn} and later, in 
\refs{\SAcurve,\KazakovDU}, where the string theory instantons were associated 
with the double points of  of the complex curve. In this paper we
study the quantum corrections to the leading  NPC to the partition function, 
given by the pre-exponential factors. We calculate these factors  
using the matrix model description of the $c=1$ string theory.

We perform the calculation by two different methods. 
The first method uses the fact that the free energy of the perturbed 
theory as a function of the cosmological constant $\mu$ and sine-Liouville 
coupling $\l$ satisfies Toda partial differential equation. The second   
method is based on the formulation of the matrix model as a system 
of free fermions and consists in direct evaluation of the quasiclassical 
fermion wave functions. This method gives the answer for a general  
time-dependent background in terms of the complex curve associated with it.

We are using the chiral formalism in which the role of canonical 
coordinate and momentum are played by the 
left and right chiral combinations $\xpm \sim x\pm p$. The advantage of this formalism is
the exact bosonization of the fermion operators. Using the
bosonization we interpret our results in terms of (target-space) CFT correlation
functions. We show that the subleading corrections are related to the
two-point correlation function of the bosonic field.
Unlike the bosonization formulae for other matrix models for
non-critical strings \refs{\KostovXI,\HanadaIM,\DijkgraafVP}, here we
have two chiral fields associated with the left and right moving
tachyons. The resulting bosonic field theory can be viewed also as a
boundary CFT, in which the boundary condition relates
the left and right fields, see {\it e.g.} \refs{\BoyarskyJB,\BoyarskySM} for
similar statement for the normal matrix model. Since our bosonic
field describes chiral excitations above the Fermi sea, its
correlation functions do not have direct interpretation in terms of
the usual FZZT branes. Nevertheless, the one-point functions,
evaluated at the singular points of the complex curve, reproduce the
disk partition function on ZZ branes, as is the case in the $c=0$
matrix model \refs{\HanadaIM,\KutasovFG}.

\newsec{Time-dependent backgrounds in Matrix Quantum Mechanics}

\subsec{The singlet sector of MQM as a system of free fermions}

\noindent
The matrix quantum mechanics can be viewed as reduction of a
two-dimensional $U(N)$ gauge theory to one dimension, and involves
one gauge field $\Ab_{ij}$ and one scalar field $\Xb_{ij}$, both
hermitian $N\times N$ matrices. The modern interpretation of the
matrix path integral is as an effective open string theory on $N$ D0
branes \refs{\McGreevyKB\MartinecKA\KlebanovKM-\McGreevyEP}. The
matrix variable $\Xb$ describes the open-string tachyon field in
presence of the $N$ D0 branes, and only effect of the gauge field is
that it projects onto the singlet sector. The theory is described
by the action
\eqn\Maction{
\CS= \int dt \Tr \( \Pb\, \nabla_\Ab \Xb - \hf (\Pb^2- \Xb^2)+\mu N \) ,
}
where $\nabla_\Ab \Xb = \p_t \Xb -i [\Ab, \Xb]$ is the covariant time
derivative.  The cosmological constant $\mu$ is introduced as a
chemical potential for the size $N$ of the matrices, which is treated
as a dynamical variable.

The closed $c=1$ string theory appears as a theory of collective
excitations of the matrix variables. The momentum modes arise as
collective excitations of the matrix $\Xb_{ij}$, while the winding
modes are collective excitations of $\Ab_{ij}$. The vertex operators
\vvintegr\ can be represented in the matrix model by\foot{When
comparing the results in CFT and matrix model descriptions, we
encounter the old problem of operator mixing \MooreIR. This problem
occurs because the correlation functions are integrated over the world
sheet and the integrals have contributions from the coinciding points
and a special prescription is needed to distinguish two tachyons with
momenta $p_1$ and $p_2$ close to each other from a single tachyon with
momentum $p_1+p_2$.  This ambiguity possibly leads to an analytic
redefinition of the couplings when passing to the matrix model
description.} \JevickiQN
\eqn\vertopM{
\CT_{p}\leftrightarrow  \cases{e^{-pt} \,\Tr \Xbp^{|p| } &
\qquad  if $p>0$ \cr e^{-pt} \,
\Tr \Xbm^{|p| } & \qquad if $p<0,$}
}
where $\Xp$ and $\Xm$ are the chiral combinations of the matrix coordinate
and momentum
\eqn\YpYm{
\Xbpm = {\Xb\pm \Pb\over \sqrt{2}}.
}

The theory simplifies significantly if formulated directly in terms of
the chiral variables $\Xbpm$ \refs{\AKK}. In the new variables the
action \Maction\ becomes
\eqn\hamfref{
\CS = \int dt \Tr \( \Xbp \nabla_\Ab \Xbm +\Xbp\Xbm+\mu N  \)
}
so that the new Hamiltonian is linear in the canonical coordinates and momenta.

In the singlet sector, characterized by absence of winding modes, the
matrix model is described by a system of free fermions whose phase
space is the spectral plane $(\xp,\xm )$ of the two commuting matrices
$\Xbp$ and $\Xbm$. The fermions are governed by a one-body
Hamiltonian
\eqn\oneph{
\hat H_0 = -\hf (\hat \xp\hat \xm +\hat \xm\hat \xp),
}
where
\eqn\ccrxx{
[\hat \xp,\hat \xm]= -i.
}
The  one-particle  wave functions  in $``\xp"$ and $``\xm"$ representations
are related by Fourier transformation $\hat S$:
\eqn\Fouriertr{\eqalign{
\psim(\xm) &= [\hat S \psip](\xm) \ \equiv \ 
\frac{1}{ \sqrt{2\pi}}\int d\xp e^{ i\xp\xm} \psip(\xp).
}}

The one-body Hamiltonian has continuous spectrum and is diagonalized,
in the $\xp$ and $\xm$ representations, by  the functions
\eqn\wavef{  \psepm(\xpm)\  = \ 
 \frac{1}{\sqrt{2\pi }}\  \xpm^{ \pm i E-{1\over 2}} \ ,
\qquad E\in {\Bbb R}.
}
The wave functions \wavef\ have  branch points at $\xpm=0$ and therefore 
are defined unambiguously only  for $\xpm>0$. It is useful to consider them  
either as multivalued meromorphic functions of the complex variables $\xpm$, 
or as analytic functions of $\log \xpm$.
 
We will restrict ourselves to the case of a theory with one Fermi sea on 
the right of the top of the potential, in which the wave functions are supported, 
up to exponentially small terms, by the positive axis. Therefore we will define  
$\psep$ and $\psem$ by \wavef\ along the positive axis, and their values along 
the negative axis will be obtained by analytic continuation from the upper and 
lower half plane, respectively.\foot{In a theory with two Fermi seas one should  
introduce a second set wave functions, which are obtained from $\psepm$ by reflection 
$\xpm \to -\xpm$. For details see the appendix of \AKK.}
The wave functions \wavef\ are orthonormal with respect to the scalar product
\eqn\sclrprdct{
(f,g)=\int _0^\infty 
d \xpm \overline{ f(\xpm)}\ g(\xpm),
}
where the integration with respect to the chiral phase space
coordinates is performed only along the positive real axis, $\xpm>0$.
The restriction of the phase space to $\xpm>0$ can be intuitively
understood as follows.  The wave function describes the system at
given moment of time and its form for other moments is obtained by
applying the evolution operator $e^{i t \hat H}$. On the other hand,
since the evolution operator shifts ${\rm arg} (\xpm)$ by $\mp it$,
the analytic continuation of the wave functions \wavef\ off the
positive axis can be viewed as the result of evolution in imaginary
time direction.

The Fourier transformation $ \hat S $ defined by \Fouriertr\ gives
the reflection part of the scattering operator, relating the incoming
leftmovers and outgoing rightmovers. It acts diagonally on the wave
functions \wavef:
\eqn\phasef{ 
\hat S \,   \psep =   e^{ i \phi_0(E)}  \, \psem\ ,
}
with
\eqn\rfacts{
e^{i\phi_0(E)} =\frac{1}{\sqrt{2 \pi}}\, 
e^{-{\pi\over 2} (E- i/2)} \, \Gamma( iE + 1/2) .
} 
The reflection phase $\phi_0(E)$, 
is the same as the one calculated using the standard
$x$-representation of the upside-down oscillator in
\refs{\MooreSF,\GinspargIS}. It has an exponentially small imaginary
part \eqn\imphas{ \Im \phi_0(-\mu)= \hf \log (1+e^{-2\pi\mu} ), }
which determines the flow of particles under the potential barrier.

\subsec{Partition function and density of states}
 
\noindent
The string theory compactified at time interval $\b = 2\pi R$ is
described by the grand canonical ensemble of fermions at finite
temperature $1/\b $ and chemical potential $\mu$. If $\CZ(\mu) $ is
the partition function of the ensemble of fermions, then the free
energy $\CF = \log\CZ$ is given by
\eqn\FRENO{
\CF\(\mu \)=  \int_{-\infty}^\infty
d E\, \rho(E )\log\left(1+e^{-\beta(\mu+E)}\right),
}
where $\rho(E )$ is the density of states related to the phase $\phi(E)$
of the fermion scattering by
\eqn\DENs{
\rho(E)= {\log \Lambda\over 2 \pi}-
{1\over 2\pi } {d\phi(E)  \over d E} .
}
The non-trivial part of the density adds to a constant cut-off
dependent term, where the cut-off $\Lambda$ is introduced as the
volume of the $(\xp,\xm)$ phase space. We remind the derivation of
this relation in Appendix A, where we also recall the derivation of
the relation between the free energy and the scattering phase
\eqn\FRENOI{
2\sin \frac{\p_\mu }{2R} \cdot \CF(\mu ) = \phi(-\mu) .
}
As we know from \KlebanovQA, this last identity allows to evaluate all
tachyon correlation functions in the compactified theory, once their
counterparts in the theory with $R=\infty$ are known.

The identities \DENs\ and \FRENOI\ hold for any Fermi system at finite
temperature and, in particular, after an arbitrary time-dependent 
perturbation that preserves the singlet sector. Thus all the information 
about the system is contained in the phase of fermionic scattering $\phi(E)$ 
as a function of the coupling constants $\{t_k\}$ associated with the
tachyon operators \vertopM.

Let us recall the explicit expressions for the stationary background,
where all $t_k=0$. Then the phase $\phi_0(E)= \phi(E)_{\{t_k=0\}}$
is given by \rfacts\ and the free energy itself is calculated by
inverting the finite-difference operator in \FRENOI. The latter is
diagonalized by the Fourier transformation
\eqn\psio{
\phi_0(-\mu)
= -{i\over 2}\, \int^{\infty}_{1\over \L} {ds\over s} 
{e^{i\mu s}\over\sinh {s\over 2} } ,
}
where we reintroduced the cut-off $\L$.  Then \FRENOI\ yields\foot{Up
to cut-off-dependent terms this integral gives the logarithm of the
Barnes \Barnes\ Gamma function: $
\CF(\mu,0)= \log \G_2( \frac{1+R}{2} -i\mu|1,R)
$.}
\eqn\invosc{
{\cal F}(\mu)_{\{t_k=0\} }=-{1\over 4}\, {
}\int^{\infty}_{{1\over \L}} {ds\over s} {e^{i\mu s}\over
\sinh {s\over 2}\,\sinh{s\over 2R}} .
}
The real part of this integral is the well-known integral
representation for the partition function of the compactified $c=1$
string theory \KlebanovQA. Its genus expansion (which is also the
expansion in $1/\mu$) reads
\eqn\FrenO{
\CF_{\rm pert}(\mu)_{\{t_k=0\} }
= - \frac{R}{2}\mu^2 \log \frac{\mu}{\Lambda} -
\frac{R + {1\over R}}{24} \log \frac{\mu}{\Lambda} +
R\sum\limits_{h=2}^\infty \mu^{2-2h} c_h(R),
}
where the genus $h$ term $c_h(R)$ is a known polynomial in ${1/ R}$.

In the Fermi system, the tachyons are represented by collective
excitations of the Fermi liquide \refs{\PolchinskiMB, \GinspargIS}.  
A general time-dependent background, corresponding to the tachyon 
deformation \PERTS\ of the world-sheet CFT, can be introduced by 
changing the state of the fermi system in the infinite past and 
in the infinite future, in the spirit of the Polchinski's derivation 
of the tree-level tachyon $S$-matrix. This amounts to
modifying the asymptotics at infinity of the fermion eigenfunctions 
\wavef\ in $\xp$ and $\xm$ representations.
The spectrum of the tachyons should be of the form \pnn\ in order to
have periodicity in $\b=2\pi R$ in the imaginary time direction.
As we know, such deformations of MQM are described by Toda hierarchy 
\refs{\DMP, \KKK, \AKK}. The commuting Toda flows can be formulated 
either as an hierarchy of PDE with respect to the `times' $t_{\pm k}$, 
or in terms of a pair of Lax operators satisfying a string equation \KostovWV. 
Below we remind the necessary facts about the two approaches.

\subsec{Time-dependent backgrounds via Toda hierarchy}
   
\noindent
The partition function of the perturbed theory is a $\tau$-function of
Toda hierarchy and satisfies a hierarchy of PDE with respect
to the couplings $t_{\pm k}$. The first of them is the Toda equation,
which is sufficient to describe sine-Liouville perturbation
$t_1=t_{-1}=\l$. Written for the free energy $\CF=\log\CZ$, this
equation has the form
\eqn\todafd{
\frac{1}{4}\,\lambda^{-1}\p_{\lambda}\lambda\p_{\lambda} \FSL(\mu, \lambda)+
\exp\left[-4\sin^2\(\frac{\p_\mu}{ 2R}\) \FSL(\mu,\lambda)\right]=1
}
and should be solved with initial condition provided by \FrenO.
  
Equation \todafd\ defines the flow between the critical points $\l=0$
and $\mu=0$ of the world-sheet CFT. When $\lambda$ is large, so that sine-Liouville
term sets the scale, the cosmological term can be considered as a
perturbation. In this case it is convenient to introduce the
following variables\foot{This definition is adjusted to the case
$R<1$; the case $R>1$ can be treated similarly.
}
\eqn\scpd{
y=\mu\xi, \qquad 
\xi=\( \l^2\, \frac{{1-R}}{R^3 } \)^{-{R\over 2R-1}}.
}
The variable $y$ is a dimensionless parameter and $\xi\sim\gst$.
Therefore, the genus expansion of the free energy is an expansion in
$\xi$ with $y$-dependent coefficients. For each coefficient in the
expansion the PDF \todafd\ yields an ordinary differential
equation \KKK. Below we will only need the genus zero contribution
$\CF_0$ to the free energy. In the variables \scpd\ it has the form
\eqn\gnexpCF{
\CF_0(\mu,\lambda)=\frac{1}{\xi^{2}} \[\frac{R}{2}y^2\log\xi+ f_0(y)\], 
}
the differential equation satisfied by    $ f_0$ can be integrated to
an algebric  equation  for the function
\eqn\Xofy{X(y)\equiv \p^2_y f_0(y),}
namely
\eqn\wf{
y=e^{-{1\over R}X(y)}-e^{-{1-R\over R^2}X(y)} .
}

\subsec{Time-dependent backgrounds via fermion wave functions}
 
\noindent
As we mentioned, in the fermionic picture one can introduce 
sources for incoming and outgoing tachyons by changing the asymptotics 
of the one-particle wave functions. The perturbed one-fermion wave functions
are obtained from the bare wave functions \wavef\ by multiplying with
a coordinate-dependent phase factor
\eqn\asswave{ 
\Pepm(\xpm)= e^{\mp i \vp_{\pm}(\xpm;E)}\,\psepm(\xpm).  
} 
The phases can be written as a sum of three terms
\eqn\pot{ 
\vp_\pm (\xpm;E)= V_\pm(\xpm) +\hf \phi(E) + v_\pm(\xpm;E),
} 
where $V_\pm$ vanishes for $\xpm=0$, $v_\pm$ vanishes for large
$\xpm\to\infty$, and $\phi(E)$ is a constant. The deformations
\PERTS\ of the world sheet CFT correspond, up to a possible analytic 
redefinition of the couplings, to the choice
\eqn\Vbig{
V_\pm(\xp)= \sum\limits_{k\ge 1} t_{\pm k} \, \xpm^{k/R}.
}
The constant mode $\phi(E)$ and the pieces $v_\pm(\xpm;E)$  
depend implicitly on $t_{\pm n}$. They
are determined from the requirement that the left and right fermions
are related by the Fourier transformation \Fouriertr:
\eqn\PepSm{
\hat S  \Pep= \Pem,\quad \hat S^{-1}  \Pem= \Pep
}
where we absorbed the reflection phase factors $e^{i\phi(E)}$ in the
definition of the deformed wave functions.
 
For large negative energy $E$, the compatibility of the
saddle-point equations for the two integrals in \PepSm\ gives
\eqn\eqfs{\eqalign{
\xp\xm&
=  \frac{1}{ R}
\sum\limits_{k\ge 1}  k t_{ k} \, \xp^{ k/R} -E  +
\frac{1}{ R}\sum\limits_{k\ge 1} v_{ k}(E)\, 
\xp^{-k/R}\cr
&
= \frac{1}{ R}
\sum\limits_{k\ge 1}  k t_{- k} \, \xm^{ k/R} -E +
\frac{1}{ R}\sum\limits_{k\ge 1} v_{- k}(E)\, \xm^{-k/R}.
}}
This means that the two  equations \eqfs\ define two functions
$\xpm=\txpm(\xmp)$ which are inverse to each other:
\eqn\invxpm{
\txp(\txm(\xp))=\xp.
}

The functions $\Xpm(\xmp)$, taken at the Fermi level $E_F=-\mu$,
can be considered as the gluing functions for an analytic curve $\CM$ 
in ${\Bbb C^2}$. This curve yields all the information for the given background.    
Its real section determines the shape of the Fermi sea in the quasiclassical 
limit $\mu\to\infty$. The functions $\Xpm(\xmp)$ are in general multi-valued, but
there is always a global parameter, the ``proper time'' variable $\t$,
such that the solution of \eqfs\ is given by $\xpm = \xpm(\t)$. If
all couplings $t_{\pm k}$ with $k>k_{\max}$ vanish, then the solution
is of the form \KostovWV
\eqn\xpmo{
\xpm(\t) =e^{\pm \t-{1\over 2R}\chi} 
\(1+ \sum\limits_{k=1}^{k_{\max}} a_{\pm k}\, e^{\mp {k\over R}\t} \),
}
where the quantity $\chi$ is related to the  constant  mode  of the phase 
\pot\  via
\eqn\suscphi{
\chi=-R \frac {\p}{\p_E}\phi.}
The coefficients $a_{\pm k}$ and $\chi$ can be found by substituting
\xpmo\ into \eqfs\ and comparing coefficients in front of $e^{\pm
{k\over R}\t}$.  For example, in the case of sine-Liouville
deformation, {\it i.e.} when only the first coupling constants 
are nonzero, this procedure gives
\eqn\param{
\mu e^{ {1\over R} \chi} -\frac{1}{R^2}
\left(1-\oR\right)\l^2 e^{{2R-1\over R^2} \chi} =1,
\qquad
 a_{\pm 1}= \frac{\l}{R} \, e^{{R-1/2\over R^2} \chi}.
}
Taking into account that $\p^2_\mu\CF_0=\chi$, it is easy to show that
this equation is equivalent to the solution \wf\ of Toda equation.
The solution beyond the tree level can
be found using the full Toda integrable structure.

The parametrization \xpmo\ can be thought of as a canonical
transformation relating the phase space coordinates $(\xm, \xp)$ and
$(\t, E)$, where the proper time $\t$ appears as the variable
canonically conjugated to the energy $E$:
\eqn\cantr{
\{\xm,\xp\}=1\  \ \Leftrightarrow \ \ \{\t, E\}=1.
}
The origin of the last relation can be traced to the Lax
representation of the operators $\xpm$ in the basis of the deformed
wave functions, where the parameter $\o= e^{\t}$ arises as the
classical limit of the shift operator $\hat \o = e^{-i \p_E}$
\refs{\AKK,\KostovWV}.

\newsec{NPC in sine-Liouville string theory from Toda equation}

\noindent
In this section we restrict ourselves to the sine-Liouville
deformation with coupling $\l$, in which case the free energy can be
determined from Toda equation \todafd. The initial condition is
provided by the free energy of the non-deformed theory which is given
by the integral \invosc. Before studying the theory with the
sine-Liouville deformation, let us remind the exact expression for
the NPC in the non-deformed $c=1$ string theory following from the
integral representation \invosc. The integral has a small imaginary
part, which describes the flow of eigenvalues beyond the top of the
inverse oscillator potential. It can be evaluated by extending the
contour to the whole real axis and taking the residues at the two
series of poles, $s_n=2i\pi n$ and $s_n=2i\pi R n$:
\eqn\noprpr{
\CF_{\rm np}(\mu)_{\{t_k=0\} }=i\sum_{n}{e^{-2\pi n \mu}
\over 4n(-1)^{n}\sin{\pi n\over R}}+
i\sum_n{e^{-2\pi R n \mu} \over 4n(-1)^{n} \sin(\pi R n )}.
}
This expression is of course compatible with \imphas\ and \FRENOI. We
see that there are two types of NPC, which have their origin in the
two kinds of branes in the $c=1$ string theory \KAK. The first type is
due to the D-instanton with Dirichlet boundary conditions on both
matter and Liouville fields, whereas the second one is due to a
D0-brane with Neumann boundary condition for the matter field.  In
both cases the pre-exponential factors do not depend on $\mu$ and
therefore scale like $\gst^0$.

The two types of NPC behave very differently with sine-Liouville
coupling $\l$.  The NPC of the type $e^{-2\pi R n \mu}$ do not flow
with $\lambda$ and are always given by the second term of \noprpr.
Indeed, given a solution of Toda equation \todafd, by adding a linear
combination of exponents $e^{-2\pi R\mu k}$ with $\l$-independent
coefficients one obtaines another solution.

In contrast, the NPC given at $\l=0$ by the first term of \noprpr,
evolve non-trivially with $\l$. The leading exponential contributions
of this series were analyzed in \refs{\KAK, \SAmn} for deformation by
vortices. From these works we know\foot{The expressions obtained in
\refs{\KAK, \SAmn} should be used after the T-duality transformation
$\xi\to \xi/R, \ y\to y.$} that the leading non-perturbative effects
have the form
\eqn\leadcor{
\eps_n(\mu,\lambda)\sim
e^{-2 g_n(y)/\xi},
}
where 
\eqn\gphi{
g_n(y)=y\theta_n(y) +\frac{1}{\sqrt{\alpha}}\,
e^{-{X(y)\over 2R^2} }\sin \frac{\theta_n(y)}{R}, \qquad
\alpha= \frac{1-R}{R(2R-1)^2}
}
and $\theta_n(y)=\partial_y g_n$ is found as a solution of the algebraic equation
\eqn\zzz{
\sin \theta_n = \(\frac{1}{R}-1\)^{-{1\over 2}} \, e^{{2R-1\over 2R^2}X(y)}
\sin \( \frac{1-R}{R}\, \theta_n\).
}
The different solutions of \zzz\ are labeled by the integer $n$ in such way that
$\theta_n\to \pi n$ when $\lambda\to 0$.

Now we would like to find the subleading order, {\it i.e.} the factor
in front of the exponential in \leadcor.  Following the procedure
suggested in \KAK, we take two solutions $\FSL$ and $\tilde \FSL$ of
\todafd\ that differ by the exponentially small quantity $\eps = \tilde \FSL-\FSL$. 
Then $\eps$ must satisfy the linearized equation
\eqn\todasph{
\frac{1}{ 4}\lambda^{-1}\p_{\lambda}\lambda\p_{\lambda}\eps(\mu, \lambda)
-4\,e^{-{1\over R^2}\p^2_{\mu}\FSL_0(\mu,\lambda)}
\sin^2\(\frac{\p_\mu}{ 2R}\)\eps(\mu,\lambda)=0,
}
where in the exponent in the second term we approximated
\eqn\sphfr{
4R^2\sin^2\(\frac{\p_\mu}{ 2R}\) \FSL(\mu,\lambda)\simeq
\p^2_{\mu}\FSL_0= R\log\xi+\XX(y).
}
This approximation is correct for our purpose since the subleading
non-perturbative contribution is of order $O(\gst)$ with respect to
the leading one, whereas the perturbative genus expansion goes in
powers of $\gst^2$. Finally, changing the variables from $(\lambda,
\mu)$ to $(\xi,y)$ we write equation \todasph\ as
\eqn\eqep{
\alpha\xi^2(y\p_y+\xi\p_\xi)^2\eps(\xi,y)
=4\, e^{-{X(y)\over R^2}}
\sin^2\(\frac{\xi}{ 2R}\,\p_y\)\eps(\xi,y).
}
This equation is solved by a refinement of the Ansatz \leadcor:
\eqn\anz{
\eps(\mu,y)\simeq\,  \, A(\xi,y)\, e^{-2g(y)/\xi}, 
\qquad
A(\xi,y)= \[\xi a(y)\]^{b },
}
where $b$ may depend on $y$.
We omit the index $n$ since it does not appear explicitly in
the equations. The details of the solution are presented in Appendix B. 
The result is that $b={\rm const}$ and $a(y)$ is given by (B.9).
 
The constant $b$, which gives the power of $\gst$ in the
pre-exponential factor, is not fixed by Toda equation. It looks
like an integration constant. Therefore, the first idea is that it
can be fixed from the initial condition \noprpr\ at $\lambda=0$,
which implies that it should vanish. On the other hand, let us
consider the limit of small $\lambda$. In this limit $y$ is large and
\eqn\corr{
\theta _n(y) \approx
\pi n+     \(\frac{1}{R}-1\)^{-{1\over 2} }
\ {\sin\frac{\pi n}{R} }\, y^{-{2R-1\over 2R}}.
}
This leads to the asymptotics
\eqn\Aass{
A(\xi,y)\sim \( {\xi / y}\)^b y^{2R-1 \over 4R}
\sim \l^{-1/2}\ \mu^{{2R-1 \over 4R}-b}  .
}
We observe that the subleading contribution does not have smooth small
$\lambda$ limit. Therefore, the parameter $b$ is not fixed by this
approach. As we will see in the next section, it should be fixed to
give the same power of $\gst$ as in $c<1$ string theories, 
\eqn\bfix{
b=\hf, 
} 
so that the full prefactor is given by the following function
\eqn\Asolf{
A(\xi,y)=C\left( e^{-\frac{1}{ R}\chi}\, \sin^2\theta \,
\sin^2\frac{ \theta }{ R}\,
\[\(\frac{1}{R}-1\)\cot\( \frac{1-R}{R}\,
\theta \)-\cot \theta\]\right)^{-1/2}.
}
The overall coefficient $C$ remains undetermined. Thus, we see that
sine-Liouville deformation changes drastically the behavior of
the non-perturbative corrections in the subleading order: the power of
$\gst \sim\xi$ changes discontinuously from 0 to 1/2. This means that
the limits $\L\to\infty$ and $\l\to 0$ do not commute. As we will see
later, the change of the behavior of the non-perturbative corrections
can be explained with the fact that the tachyon perturbations break
the time translation symmetry of the theory.

A nice consistency check of the result \Asolf\ is to reproduce the
non-perturbative corrections to the free energy of the pure gravity in
the limit where the couplings approach the $c=0$ critical point. We
discuss this limit in Appendix C where we show that the result \Asolf\
does reproduce the correct critical behavior.

\newsec{NPC from the quasiclassical wave functions}

\noindent
The approach based on Toda equation is conceptually straightforward
but does not allow to fix the solution completely. Alternatively one
can exploit the fermionic formulation to evaluate the non-perturbative
corrections to the fermion reflection phase and then use the relation
\FRENOI\ to find those for the partition function. Moreover, this
second approach allows to generalize the results obtained in the
previous section to the case of a deformation with arbitrary number
of non-vanishing couplings. Before considering the general case, we
will illustrate the method for the stationary background,
where the exact form of the non-perturbative corrections is known.

\subsec{The case of a stationary background}
 
 \noindent
Although the exact expression for the reflection phase is known, we
follow here another approach which works only quasiclassically but can
be generalized to other situations. For this we use the following
matrix element of the scattering operator \phasef
\eqn\dblint{
\frac{1}{\sqrt{2\pi}}\tint_{\!\! 0}^\infty \, d\xp d\xm\, 
\overline{\psem(\xm)}\, e^{i\xp\xm}\,\pseep(\xp)
=e^{i\phi_0(E)}\delta(E-E'),
}
where we used the orthonormality of the wave functions \wavef.
Substituting the explicit expression for the wave functions \wavef\
and introducing a cut-off $\Lambda$ equal to the volume of the phase
space, we write the diagonal matrix element as
\eqn\dblcut{
\frac{1}{ (2\pi)^{3/2}}\tint _{\!\! 0}^{\sqrt{\Lambda}} \
\frac{d\xp d\xm }{ \sqrt{\xp\xm}}\, e^{i\(\xp\xm+E\log(\xp\xm)\)}
= e^{i\phi_0(E)}\rho_0(E),
}
where $\rho_0(E)\approx {1\over 2\pi} \log(-\L/ E)$ is the density of
states corresponding to the cut-off $\L$. In the following discussion
we can safely approximate $\rho_0(E)$ by $\frac{1}{ 2\pi}\log\Lambda$.

We would like to evaluate the integral \dblcut\ by the saddle point
method. The two saddle point equations associated with the variables
$\xp$ and $\xm$ actually coincide and both give the equation for the
classical fermion phase-space trajectory with energy $E$:
\eqn\nonfs{
\xp\xm=-E.
}
The fact that the two equations coincide means that we do not have
isolated saddle points but `saddle contours' that go along the flat
direction. If we change the variables to
\eqn\xpmonon{
\xpm(\tau)= \sqrt{\e}\, e^{\pm \tau},
}
then the flat direction is along the `proper time' $\t$.  The
cut-off prescription $0<\xpm<\sqrt{\L}$ then restricts the
$\tau$-integration in \dblcut\ to the interval
\eqn\cutoff{
-\hf \log\frac{\Lambda}{\e }<\tau<\hf \log\frac{\Lambda}{\e }
}
and the integral in $\t$ gives the factor 
$ \log {\L\over \e} \approx 2\pi \rho_0(-\e)$.
The remaining integral in $\e$ is 
\eqn\dblnew{
\int _0^\L\frac{d\e}{\sqrt{2\pi \e}} \,
e^{i(E\log\e+\e )} = e^{i\phi_0(E)} .
}
The saddle point is at $\e=-E$ and the gaussian fluctuations cancel
$\sqrt{2\pi\e }$ in the denominator. This gives the genus-zero
perturbative contribution for the reflection phase:
\eqn\refph{
\phi_0(E)\approx E\log (-E)-E .
}

The appearance of the non-perturbative corrections is related to the 
fact that the phase of the integrand in \dblcut\ is a multivalued function
of $\e$.  As a consequence, there is an infinite number of saddles at
\eqn\logepsn{
\log\e_n = \log(-E) - 2\pi i n 
}
with $n$ integer. The saddles with $n>0$ can be connected by
constant phase contours to the dominant saddle and thus are
potentially relevant. The integral along a contour that passes
through the $n$-th subdominant saddle contributes a factor $(-1)^n
e^{2\pi n E} \, e^{i\(E\log (-E)-E\)}$, where $(-1)^n$ comes from the
pre-factor in \dblnew.  Taking into account the contribution of all
saddles, one finds for the Fermi level $ E=-\mu$
\eqn\refphnnn{
e^{i\phi_0(-\mu)}\approx e^{-i\(\mu\log \mu -\mu\)}
\(1 + \sum_{n\in {\Bbb N} } c_n \, (-1)^n  \,e^{-2\pi n \mu}\),
}
where the coefficients $c_n$ depend on the choice of the integration contour. 
This is in agreement with the expression
that follows from the exact answer \imphas,
\eqn\refphnn{
\phi_0(-\mu )\approx -\mu \log \mu +\mu   -i\,
\sum_n {\frac{1}{2n}\ (-1)^n}\,e^{-2\pi n\mu} .
} 
The coefficient $c_1=\hf$ in the contribution of the first subdominant
saddle can be explained by the fact that the contour turns at $\pi/2$
after reaching the saddle \CallanPT.  
The choice of the integration contours and the evaluation of the contributions 
of the subdominant saddles is a delicate problem and we will not discuss it here.

Instead, we will try to understand the geometrical meaning of the saddles
\logepsn\ in terms of the complex curve defined by the saddle point equation \nonfs\    
at the Fermi level $E_F=-\mu$,
\eqn\ccmo{
\xp\xm = \mu.
}
As a real manifold the curve represents a hyperboloid, depicted in fig. 1a.   
It is characterized by two non-contractible cycles, the compact $A$-cycle, 
given by the section $\bar \xp = \xm$, and the non-compact $B$-cycle given by   
the connected component of the section $\bar \xpm = \xpm$ with $\xpm >0$, 
which coincides with the profile of the Fermi sea. The $B$-cycle connects 
the two infinite points $\infty_-=\{\xm=\infty\}$ with $\infty_+=\{\xp=\infty\} $.
A global parametrization of $\CM_0$ is given by the map
\eqn\globl{
\xpm = \sqrt{\mu }\, e^{\pm \t}
}
to the strip $ |\Im \t | \le \pi$. 

\figtwo{\ninepoint The Riemann surface of the non-deformed theory (one
dimension is suppressed). The ``saddle contours" are given by the 
$B$-cycle and $n$ copies of the $A$-cycle. 
The second picture represents the covering of the hyperboloid by 
the complex $\tau$-plane and non-minimal saddle contours for $\xpm(\tau)$ 
in the universal cover.}{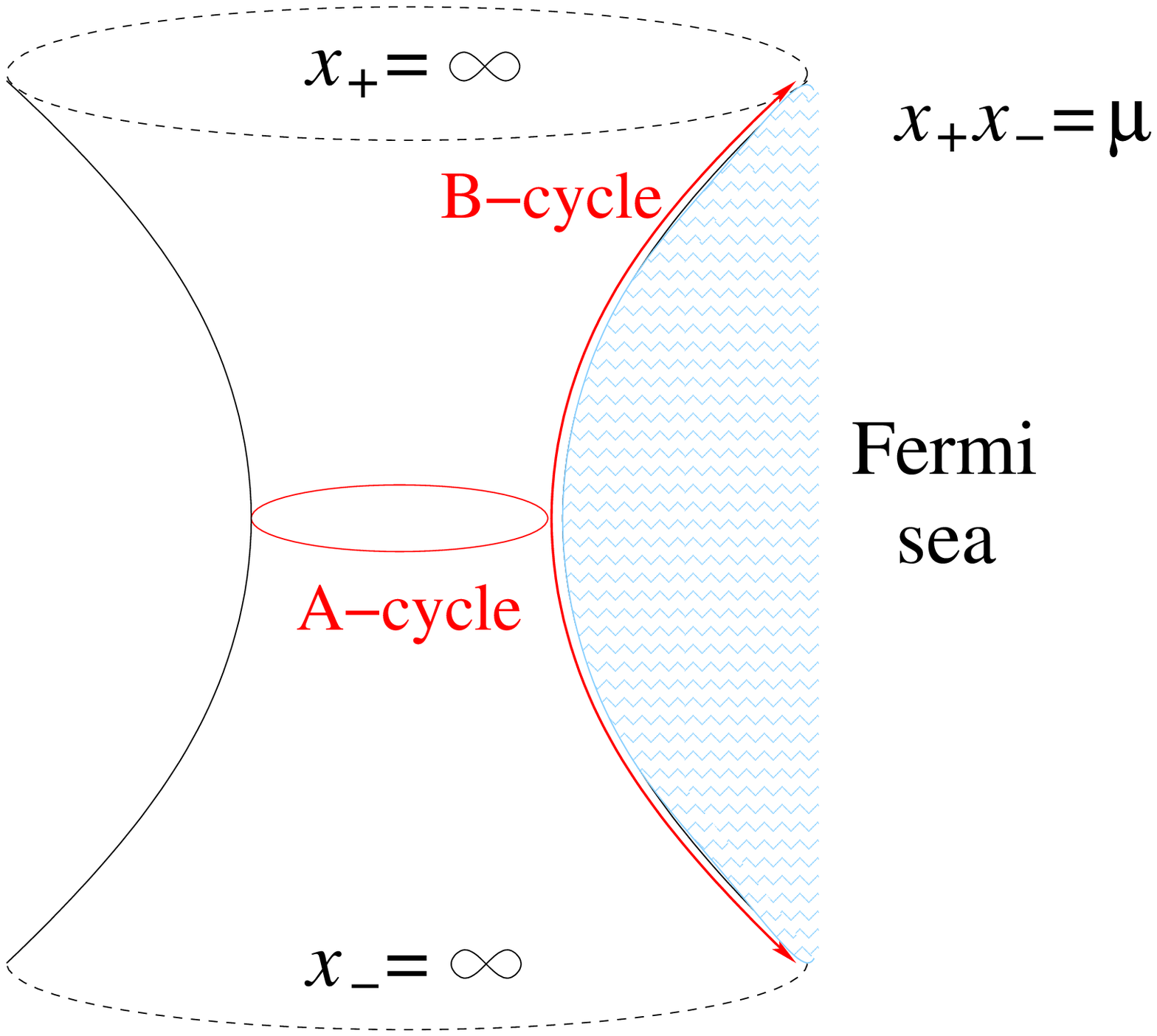}{5cm}{0.5cm}{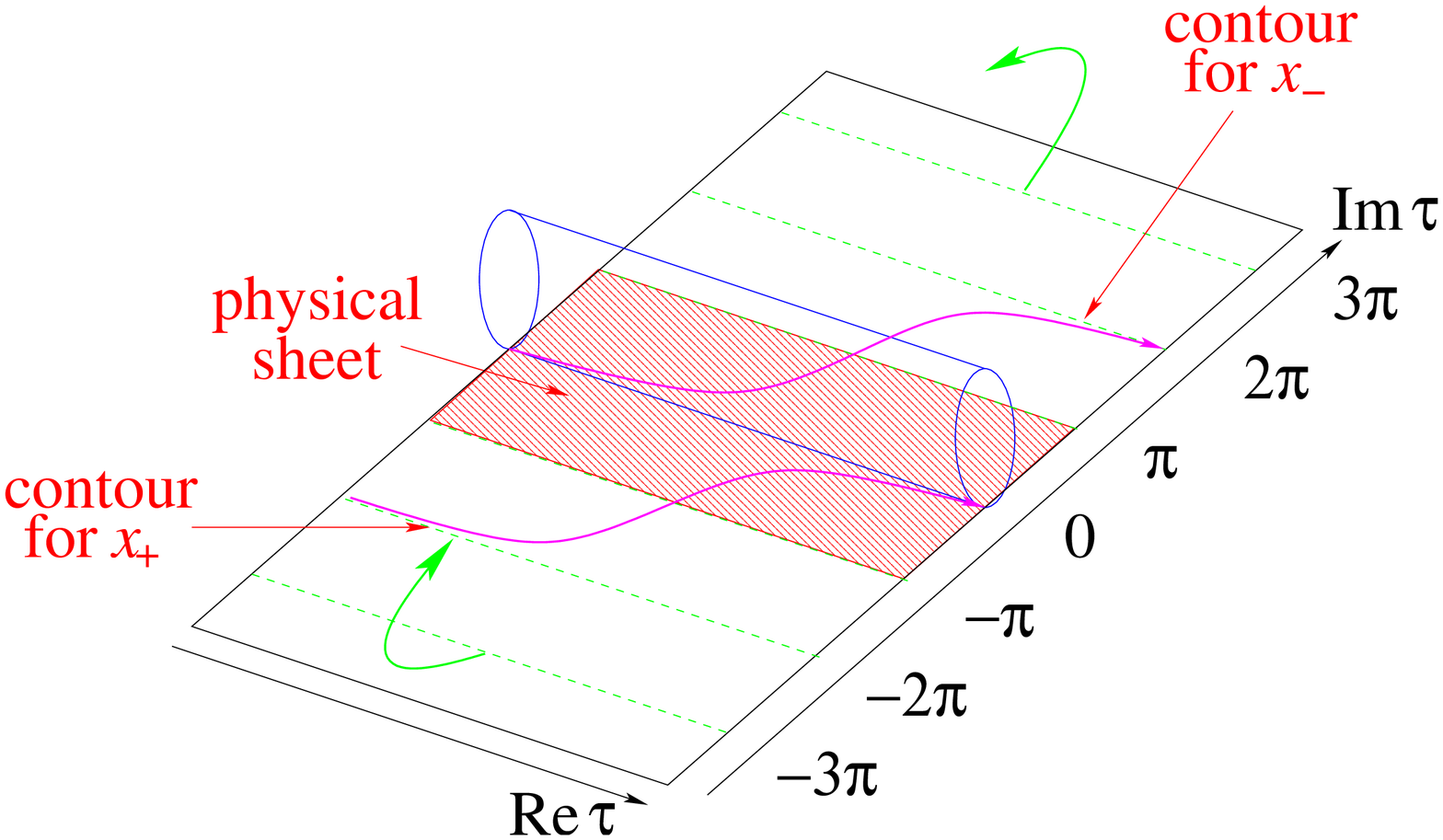}{7.5cm}  

However, because of the multi-valuedness of the integrand in \dblcut,
the solutions of the saddle point equation are actually described by
the universal cover $\overline{\CM_0}$ of the curve \ccmo, globally
parameterized by the whole complex $\tau$-plane. 
For the minimal saddle, $n=0$, the integration contours for $d\xp$
and $d\xp$ belong to the same sheet of the universal cover $\CM_0$
and coincide with the cycle $B$ on the hyperboloid, which is
parameterized by the real axis in the $\t$-plane.

In contrast, the non-minimal saddles, $n\ge 1$, are described by two {\it different}  
contours on the universal cover $ \overline{\CM_0}$:
one of the contours describes $\xp(\tau)$ and the other one gives
$\xm(\tau)$. In the parametrization \globl, they are represented as
contours joining $-\infty -2\pi i n$ to $\infty$ and $-\infty$ to
$\infty+2\pi i n$, respectively, as the ones drown in fig. 1b. These
conditions ensure that $\Im(\log \xpm) =0$ in the asymptotic regions
of large $\xpm$. The two contours have the same projection in
${\CM_0}$ where they follow the cycle $B$ and at some point wind $n$
times around the cycle $A$, as it is shown on fig 1a.
But since $\xp$ and $\xm$ follow the different contours in the universal cover,
the integrand in \dblcut\ along such a contour yields an additional factor  
$ (-1)^{n} e^{- 2\pi n\mu}$.
 
In this way, the non-minimal saddles are assiciated with `double
contours' on the hyperboloid \ccmo, rather than with double points. The
double points appear in the complex curve for the resolvent, $y= w(x)$,
which is also the $c\to 1$ limit
of the FZZT curves of minimal string theories.
The curve for the resolvent is parametrized as
\eqn\freepxW{
x=   \sqrt{2\mu}  \cosh \t,\qquad 
w= -  {\sqrt{2\mu} \over \pi} \,
\t\,\sinh\t
}
and represents a ${\Bbb Z}_2$ orbifold of the universal cover of the
hyperboloid \ccmo, obtained by identifying the points $\t$ and $-\t$.
There are two points on the curve ($\pm \sqrt{2\mu},0$), 
which are also the positions of the branch points of $y= w(x)$, which
are images of infinitely many points on the $\t$-plane $\t=i\pi n$.
They appear as the (degenerate) limit of the double points of the minimal string curves.

\subsec{Quasiclassical wave functions in the general case}
 
\noindent
Now we will apply the same method to calculate the NPC to the
scattering phase in the case of a general tachyon deformation
described in section 2.4. The first step of the calculation consists
in the evaluation of the quasiclassical asymptotics of the fermion
wave functions.
   
First of all, we need the classical limit of the phases
$\vp_{\pm}(\xpm, E)$ in \asswave, which is determined by the
compatibility of two equations \eqfs. The integration of these
equations gives the following representation
\eqn\intphase{
\vp_\pm(\xpm)=\int^{\xpm}_{\infty}\txmp (\pxpm)
\,d\pxpm+E\log\xpm+\phi_\pm ,
}
where the functions $\xpm = \txpm(\xmp)$ are the classical fermion
trajectories defined in parametric form by \xpmo\ and $\phi_\pm$ are
integration constants. As a result, in this approximation the wave
functions take the form
\eqn\quaswf{
\Pepm(\xpm)\approx B_{_{\pm}}(\xpm)\,{e^{\mp i \phi_\pm}}
e^{\mp i \int^{\xpm}_{\infty}\txmp(\pxpm)\,d\pxpm} ,
}
where $B_{_{\pm}}$ are factors coming from the subleading order in the
Planck constant. These factors and the zero mode $\phi=\phi_+
+\phi_-$ can be fixed from the normalization condition similar to
\dblint.\foot{Note that the deformed wave functions 
are not orthogonal anymore.}
  
With the cut-off $\Lambda$ introduced as in the previous
subsection, the wave functions satisfy the normalization condition
\eqn\normone{
\int _0^{^{\sqrt{\Lambda}}} \!\!\!\! d\xpm\,
\overline{\Pepm(\xpm)}\ \Pepm(\xpm)=\frac{1}{ 2\pi}\log\Lambda.
}
A second condition follows from the action of the scattering operator,
that is the Fourier transformation. Since we absorbed the scattering
phase into the wave functions, the Fourier transformation acts as the
identity operator, which leads, together with \normone, to
\eqn\normtw{
\frac{1}{ \sqrt{2\pi}}\int_0^{^{\sqrt{\Lambda}}} \!\!\!\! d\xp\,
\int_0^{\sqrt{^{\Lambda}}} \!\!\!\! d\xm\,
\overline{\Pem(\xm)}\, e^{i\xp\xm}\, \Pep(\xp)=\rho(E) .
}
Substituting the asymptotic form \quaswf\ and keeping only the constant
cut-off dependent piece of the density \DENs, one writes the two
conditions as
\eqn\condB{
\int_0 ^{^{\sqrt{\Lambda}}} \!\!\!\! d\xpm\,
|B_{_{\pm}}|^2={1\over 2\pi}\log\Lambda,
}
\eqn\normtwo{
\int_0^{^{\sqrt{\Lambda}}} \!\!\!\! d\xp\,
\int_0^{\sqrt{^{\Lambda}}} \!\!\!\! d\xm\,
{\bar B_{_{-}}}B_{_{+}}\, e^{iS(\xp,\xm)}
={e^{i\phi}\over \sqrt{2\pi}}\log\Lambda ,
}
where we introduced the effective action
\eqn\actxpm{
S(\xp,\xm)=\xp\xm- \int^{\xp}_{\infty}\!\!\!\! \txm(\pxp)\,d\pxp
-\int^{\xm}_{\infty}\!\!\!\! \txp(\pxm)\,d\pxm.
}

The integral in the second condition can be evaluated by the saddle
point method. The property \invxpm, following from the compatibility
of two equations \eqfs, means that the leading contribution to the
integral is again associated with a one-dimensional saddle contour
$\gf$ in the $(\xp,\xm)$ plane, defined by the functions
$\xpm=\txpm(\xmp)$ and going from $\xm=\infty$ to $\xp=\infty$. For
$E=-\mu$ the saddle contour  defines the profile of the Fermi sea.

To evaluate the gaussian fluctuations in the transversal direction to
the saddle contour, it is convenient to change variables from $(\xp,
\xm) $ to $(\t,-\e)$, where $\t$ is the same as in \xpmo\ and the
variable $\e$ parameterizes the transversal direction. The equation
of the saddle contour in the new variables is $\e=-E$. For the
quadratic form $\delta_{\e}^2 S(\xp,\xm)$ one finds, using \cantr,
\eqn\fluc{\eqalign{
\delta_{\e }^2 S(\xp,\xm)=&\(2\p_{\e } \xp \p_{\e } \xm-
\frac{d\xm}{  d\xp} \(\p_\e \xp\)^2-\frac{d\xp}{ d\xm} \(\p_\e \xm\)^2\)
\frac{(\delta\e )^2}{2}\cr
=& \(\frac{\p_{\e }\xp}{\p_{\t}\xp}- \frac{\p_{\e }\xm}{ \p_{\t}\xm}\)
\frac{(\delta\e )^2}{2}=-\frac{1}{\p_{\t}\xp \p_{\t}\xm}\,
\frac{(\delta\e )^2}{ 2}.
}}
As a result, the condition \normtwo\ is written as an integral along
the saddle contour $\gf$:
\eqn\intcondB{
e^{i\int_{\gf}\!\! \xm d\xp}\int_{\gf} \!\! d\t\, \,
\bar B_{_{-}}B_{_{+}}\,
\[ -\p_{\t}\xp \p_{\t}\xm \]^{1/2}={e^{i\phi  }\over 2\pi}\log\Lambda.
}
The two conditions \condB\ and \intcondB\ are satisfied  if  
\eqn\phiint{
\phi = \phi_{\rm pert}  (E)=\int_{\gf}\!\! \xm\,d\xp ,\qquad B_{_{\pm}}
={1\over \sqrt{\pm 2\pi \p_{\t}\xpm}}.
}
In particular, taking into account the relation \FRENOI, this result
gives the expression of the $\mu$-derivative of the perturbative free
energy $\CF_{\rm pert}$ as an integral over the $B$-cycle on the
complex curve \xpmo\ \AKKNMM.
Knowing the wave functions \quaswf, one can also reconstruct the
quasiclassical wave function in the $x$-representation. 
We refer to Appendix D for details.

\subsec{Instantons as double points of the complex curve}

\noindent
In order to find the non-perturbative corrections for the free energy
it is sufficient to calculate those for the scattering phase
$\phi(E)$.  This can be done by taking into account also the
non-minimal saddles of the integral \normtwo.
We saw in section 4.1 that if there is no tachyon potential, the
non-minimal saddles are given by pairs of one-dimensional contours
living on different sheets of the universal cover of the hyperboloid
$\xp\xm=\mu$, but having the same image in ${\Bbb C^2}$. The
collective coordinate along these saddles is the Minkowski time, since
the background is time-independent. In presence of tachyon potential
there is no time translational invariance. As a consequence, the
minimal saddle contour will evolve with time and the non-minimal
saddles will be given by isolated saddle points, or double points of the
deformed complex curve, to be described below. The meaning of these
double points is essentially the same as in the minimal string
theories \refs{\SeibergNM,\KazakovDU}.

Again, the saddle-point equations $\xp = \Xp(\xm)$ and $\xm = \Xm(\xp)$ 
define a complex curve $\CM$ in ${\Bbb C^2}$, which is a
deformation of the universal cover of the hyperboloid $\xp\xm=\mu$.
The Riemann surfaces of the functions $ \Xp $ and $\Xm $ are the
projections of the curve $\CM$ to the $\xm$ and $\xp$ complex planes.
Due to the multi-valuedness of the functions $\Xp(\xm)$ and
$\Xm(\xp)$, equation \invxpm \ has also particular solutions
describing isolated double points, where the complex curve touches
itself \SAcurve.
 
The double points can be classified most easily by using the
uniformization parameter $\t$, defined in equation \xpmo. They
represent pairs of points $\tau'\neq \tau''$ in the complex $\t$-plane
such that
\eqn\twoo{
\xp({\tau'})=\xp({\tau''}) \quad {\rm and}\quad \xm({\tau'})=\xm({\tau''}).
}
Each double point is  by construction a solution of \invxpm\ and as
such represents a saddle point for the action \actxpm.  
 
Let us restrict ourselves to the case $t_k=t_{-k}$ for all $k$, which
is simpler to analyze. Then $a_k=a_{-k}$ in \xpmo\ and the set of
double points is given by the pairs 
$(\t'= -i\theta _n,\t''= i\theta_{n})$, where $\theta _n$ are determined by
\eqn\thetak{
\sin \( \theta _n\) =\sum\limits_{k=1}^{k_{\max}}
a_k\sin \( \frac{k-R}{ R}\, \theta _n\),
\qquad \th_n \to \pi n \ {\rm when \ all }\ t_k\to 0,
}
As $\tau'$ and $\t ''$ are pure imaginary and complex conjugated,
$\xp(\pm i\th_n)=\xm(\pm i\th_n)$ are both real. For the
sine-Liouville deformation the discrete set of parameters $\th_n$
is determined by the equation
\eqn\phik{
\sin \( \th_n\) = a(y)\sin \( \frac{1-R}{R}\, \th_n\)
}
with $a(y)$ from \param. This equation is the same as \zzz\ and
therefore the parameters $\th_n$ coincide with those defined in
section 3.

Let us calculate the contribution of the $n$-th saddle point to the
normalization integral \normtwo. The leading contribution is given by
the value of the action \actxpm, which depends on the integration
contours for $d\xp$ and $d\xm$ along the complex curve. The two
contours can be parameterized in the $\tau$-plane by the intervals
\eqn\paths{\eqalign{
\t\in& \, (\infty,0)\cup (0,-i\th_n)  \qquad {\rm for\ }\xp(\t) , \cr
\t\in& \,  ( -\infty,0)\cup (0,i \th_n) \qquad   {\rm for\ }\xm(\t).
}}
As a result, the action is given by $S_{\rm pert}+iS_n$,
\eqn\contgk{
S_{\rm pert}=\int_{\gf}\!\!\xm\, d\xp, \qquad
S_n = i\oint_{\gamma_n}\!\!\xm\, d\xp,
}
where the contours $\gf$ and $\gamma_n$ are respectively  
the images of the intervals $(-\infty,\infty)$ and
$(i\th_n,-i\th_n)$ in the $\t$-plane.
 
The first term $S_{\rm pert}$ gives, according to eq. \phiint, the
leading perturbative approximation to the scattering phase. Its
integration contour is the $B$-cycle connecting the two punctures at
$\xpm = \infty$.
On the other hand, the image of the integration contour $\gamma_n$ for
the second term $S_n$ is a closed loop in ${\Bbb C}^2$, because it
ends at the $n$-th double point. If the double point is considered as
a vanishing $A$-cycle of the complex curve, then the contour $\g_n$ is
a the dual compact $B$-cycle.
 
To find the subleading pre-exponential factors, one should evaluate
the fluctuations around the saddle points. The variation of the
action \actxpm\ gives
\eqn\varph{
\delta^2 S_n(\xp,\xm)=\delta \xp \delta \xm-
\hf \left. \frac{d\xm}{ d\xp}\right|_{-i\th_n}\!\!\!\! \(\delta\xp\)^2-
\hf \left.\frac{d\xp}{d\xm}\right|_{i\th_n} \!\!\!\!   \(\delta\xm\)^2.
}
This leads to the following prefactor in the calculation of the integral \normtwo
\eqn\prefac{
\Delta_n=\left| \ \(\frac{\p\xp}{\p \tau}\)_{-i\th_n}
\(\frac{\p\xm}{\p\tau}\)_{i\th_n}
\(1- \(\frac{d\xm}{ d\xp}\)_{-i\th_n} \!\!
\(\frac{d\xp}{ d\xm}\)_{i\th_n}\) \ \right|^{-1/2},
}
where the first two factors come from the quasiclassical wave functions.

Adding together contributions to the left hand side of \normtwo\ of
the minimal saddle contour, given in \intcondB, and all isolated
saddle points, we get\foot{Here we do not specify 
the contours of integration and therefore the
combinatorial factors in front of the individual terms. We assume that 
the contribution of each saddle point enters with the factor 1/2.}
\eqn\condkkk{
e^{i S_{\rm pert}}
\[\log\Lambda  + \sum\limits_{n=1}^{\infty}
\sqrt{\pi \over 2}\,\Delta_n\, e^{-S_n}\]=e^{i \phi(E)}\,\log \Lambda.
}
Taking the logarithm of both sides, one gets a quasiclassical
expression for the zero mode that includes the non-perturbative
corrections
\eqn\phikkk{
\phi(E)\approx \int_{\gf}\!\! \xm\,d\xp
-i\sum\limits_{n>0}{\sqrt{\pi}\,\Delta_n\over\sqrt{2}\, \log\Lambda}\, e^{-S_n}.
}
Finally, we can apply \FRENOI\ to evaluate the non-perturbative terms
for the free energy itself which gives\eqn\freekk{
\CF\approx
\CF_{\rm pert}+{i\sqrt{\pi}\over 2\sqrt{2}\log\Lambda }\sum\limits_{n>0}
{\Delta_n\over \sin {\p_\mu S_n \over 2 R} }\, e^{-S_n}.
}
The classical action $S_n$ associated with  the $n$-th double
point can be written as
\eqn\Skk{
S_n=i\int\limits_{i\th_n}^{-i\theta _n} \xm \p_{\tau}\xp d\tau
=-2 \int\limits^{E} \th_n d\e,
}
where we used the canonical transformation \cantr, and the
prefactor to the exponent has the form
\eqn\Akk{
A_n \sim {\Delta_n\over \sin {\p_\mu S_n \over 2 R} }
= \(\sin^2\frac{\theta }{ R} \,
\[\(\frac{\p\xp}{\p \t}\)_{-i\th_n}  \(\frac{\p\xm}{\p \t}\)_{i\theta _n}-
\(\frac{\p\xp}{\p \t}\)_{i\theta_n}\(\frac{\p\xm}{\p \t}\)_{-i\theta_n }\]
\)^{-1/2} .
}

The free energy is determined by \FRENOI\ up to terms of the form $e^{-2\pi R\mu}$.  
The analysis of section 3 based on Toda equation shows that these
corrections do not depend on the tachyon potential and therefore are
given by the last term on the r.h.s. of \noprpr. However there is an argument,
advanced in \KAK, that such terms do not appear whenever a tachyon
potential is switched on.

The result \Akk\ gives the instanton corrections for an arbitrary
tachyon deformation of the theory. Now let us restrict ourselves to
the case of sine-Liouville deformation. It was already shown in
\SAcurve\ that in this case $S_n$ reproduce the leading
non-perturbative corrections \leadcor\ obtained by Toda equation.
Therefore let us concentrate
on the subleading contribution \Akk. Calculating the derivatives
using \xpmo, where only terms with $k=1$ are present, and taking into
account the defining equation \phik\ for $\theta = \theta_n$, one finds
\eqn\Afsolff{
A(\xi,y)=C\left\{e^{-\frac{1}{R}\chi} \sin^2 \theta \, \sin^2 \frac{\theta }{R}\,
\(\(\frac{1}{R}-1\)\cot\( \frac{1-R}{ R}\,\theta \)-\cot \theta \)\right\}^{-1/2},
}
where the overall coefficient $C$ is found to be
\eqn\coefCCC{
C={i\sqrt{\pi R}\over 4\sqrt{2}\log\Lambda }.
}
This result is in complete agreement with the result \Asolf\ obtained
by integrating Toda equation.

In particular, this confirms our claim that for any finite tachyon deformation 
the subleading corrections
scale  like $\gst^{1/2}$  and thus they are 
non-analytic in the limit $\lambda\to 0$. This is due to the breakdown of the 
time translation invariance  in any time-dependent background.
As a consequence, in the first case the contributions
to the saddle point approximation come from one-dimensional saddle contours,
whereas in the second case they arise from isolated saddle points.
Hence, the determinants of fluctuations, giving the main non-trivial
contribution to the subleading correction, are one- and
two-dimensional, correspondingly. This explains the difference in the
power of the string coupling.

\newsec{The instanton effects from bosonization}

\noindent
In the matrix model, the tachyon modes are collective excitations of
fermions propagating as left and right moving waves on the Fermi
surface. After second quantization, these modes form a bosonic
field. In $(x,p)$ representation of MQM, the fermions are
non-relativistic and as a consequence the bosonic field is
self-interacting.  In contrast, as in $(\xp,\xm)$ representation
the one-particle fermionic Hamiltonian becomes first order, the
fermions can be exactly bosonized. This bosonization is essentially
the one that occurs in Toda hierarchy, modulo some subtleties related
to the fact that we are using a non-compact realization of the latter.

In this section we show that the non-perturbative corrections to the
free energy can be expressed in terms of vertex operators for a 
boson field. The bosonic field in question describes not the whole spectrum 
of tachyons in the theory, but only  the discrete subset that survives 
after the compactification. The bosonic field formalism  works particularly 
well in the quasiclassical limit and gives a nice  interpretation to the 
results obtained in the previous two sections. It might also help to 
interpret these results  in terms of D-branes in $c=1$ string theory with 
time-dependent background.

\subsec{Second quantized fermions and density operator}

\noindent
We start  by  introducing the second quantized fermion fields 
\eqn\modepm{\eqalign{
\hat\Pspm(\xpm, t)&= \int dE \, e^{\mp  Et/2}
\, \Pepm\(e^{\mp t}\xpm\) \, b(E)  ,\cr
\hat\Pspm ^{\dagger}(\xpm, t)&= \int dE \, e^{\mp Et/2}
\, \overline{ \Pepm\(e^{\mp t}\xpm\)}\, b^{\dagger} (E) ,
}}
where the operator amplitudes $b(E)$ and $b^{\dagger}(E)$ satisfy the
canonical anti-commutation relations
\eqn\bpbmc{
\{ b^{\dagger}(E) , b(E') \} =  \d(E-E'),
}
with all other anticommutators equal to zero.
We work with only one set of operator amplitudes, 
since the left and right fermion operators are related by Fourier 
transformation and the fermion wave functions have the property 
\PepSm. The last property followed from the fact that we absorbed the 
reflection phase in the definition of the wave functions, so that 
the fermion reflection operator  acts as the identity operator 
on the amplitudes in $E$-representation.
  
We are interested in operators in the matrix model of the form
\eqn\operf{
\CO_f= \Tr f(\Xbp, \Xbm),
}
where $f(\xp,\xm)$ is a  smooth function of its two variables.
The operator $ \CO_f$ is well defined as the matrices $\Xbp$ 
and $\Xbm$ commute due to the gauge field.
In the second quantized formalism, this operator translates into 
\eqn\operUU{
\hat\CO _f= \int d\xp d\xm f(\xp, \xm) \, \hat\CU(\xp, \xm),
}
where
\eqn\densU{
\hat\CU(\xp,\xm, t) = \textstyle{1\over \sqrt{2\pi}}\, 
e^{i \xp\xm}\hat\Psm^{\dagger}(\xm ,t) \hat\Psp (\xp,t)
}
is the fermion phase space density operator.  
The expectation value of the operator $\hat\CU$, or the Wigner's function,  
is evaluated with respect to the thermal vacuum of the compactified 
theory which is defined by \KlebanovQA
\eqn\thermalv{
\< \mu|   b^{\dagger}(E)\, b(E')|\mu\> ={ \delta(E-E') \over 1+ e^{\b(\mu+E)}}.
}
In particular, for the trace of the identity operator $N\equiv \< \mu
|\Tr {1}|\mu\>$ we find, using \thermalv\ and \normtw,
\eqn\trU{\eqalign{   
N=\int d\xp d\xm \langle  \mu| \hat\CU(\xp,\xm)|\mu \rangle
&= \int _{-\infty}^\infty {dE\, \rho(E)
\over  1+ e^{\b(\mu+E)}}  .
}}
Comparing with \FRENO, one reproduces the well known relation between
the number of fermions and the grand canonical free energy \eqn\trf{
N=  -{\mu \over {2\pi}}  \log \L-{1\over {2\pi R}} \p_\mu\CF, 
}
where we explicitly included the non-universal cut-off dependent term.
Then \FRENOI\ allows also to write the relation of $N$ to the reflection phase
\eqn\sintr{
\sin \(\frac{1}{2R}\p_\mu\) N= -{1\over 4\pi R}\(\log\Lambda+   \p_\mu\phi(-\mu)\) .
}

\subsec{Bosonization formula for the fermion wave functions and the density operator}

\noindent
The fermionic operators can be expressed as exponents of a bosonic
field with continuous spectrum of energies.  It is however technically
more advantageous to introduce another bosonic field with discrete
spectrum associated with the possible energies in the Euclidean
compactified theory.  If we restrict the spectrum of tachyons to be
discrete as in the compactified theory, the spectrum of the bosonic
fields will be also restricted to the discrete set of purely imaginary
momenta $p_n = i {n/R}, \ n\in {\Bbb Z}$.  The sector of the theory
spanned on these states gives a `non-compact' realization of Toda
integrable structure.\foot{Usually in  Toda hierarchy the wave functions 
are bi-orthogonal polynomials and the $\tau$-functions are labeled by 
a non-negative integer $s\in {\Bbb Z}_+$, the degree of the polynomial \refs{\UT,
\Takasak}.  In our case the role of $s$ is played by the discrete 
complex variable $\mu + i(n+{1\over 2}){1\over R}$, $n\in {\Bbb Z}_+$.  
We use the terms compact and non-compact Toda hierarchy in the analogy 
with the representations of the compact $su(2)$ and the non-compact 
$ sl(2,{\Bbb R})$.} The bosonic field is thus spanned 
on the operators $\{E, \p_E$, $t_{\pm n},\p_{t_{\pm n}}\}$ and has the form
\eqn\smallBf{
\hat \Phpm (\xpm) = V_\pm(\xpm)+\hf\hat\phi
-E \log\xpm + \hat \Dpm (\xpm),
}
where $V_\pm$ are the potentials \Vbig\  
and the differential operators $\hat\phi$ and 
$\hat \Dpm$ are defined by
\eqn\defD{\hat\phi=- \frac{1}{R} \p_E, \qquad 
\hat \Dpm(\xpm) = \sum_{n\ge 1} \textstyle{1\over n} \,
\xpm^{-{n/R}}\,{\p_{t_{\pm n}}}.
}

The deformed fermion wave function at level $E=-\mu$, known in the
mathematical literature as Baker-Akhieser function, can be written as
the expectation value of normal ordered exponential of a bosonic
field. A straightforward generalization of the bosonization formula
for the compact Toda hierarchy  gives \KostovTK
\eqn\Boso{
\Pspm^{-\mu}(\xpm)=(  2\pi\, \xpm)^{-1/2} \, 
\< \mu|:{e^{ \mp  i  \hat\Phpm(\xpm) }}:|\mu \>,
}
where the normal product sign $: \ :$ means that all derivatives are
moved to the right, and the expectation value of the differential
operator $\hat\CO$ is defined as $\langle \hat\CO\rangle
=\CZ^{-1}\cdot \hat \CO \cdot \CZ$.  From the bosonic representation
of the Becher-Akhieser function we get the following operator formula
for the fermion bilinears
\eqn\Bilin{ 
\hat\Psm^{\dagger}(\xm) \hat\Psp(\xp)
= \frac{1}{  2\pi }  (\xp \xm )^{-\half}\,  \(:e^{- i\hat \Phm(\xm)  }:\)^{\dagger}  
:e^{- i\hat \Php(\xp)  }: \, ,
}
where the hermitian conjugation changes the normal to anti-normal ordering
in the first factor. Passing to the normal ordering and subtracting a
divergent term in the exponent, we finally obtain 
\eqn\Bilin{ 
\hat\Psm^{\dagger}(\xm) \hat\Psp(\xp)
= \frac{1}{ 2\pi R} \, (\xp\xm)^{-{R+1\over 2 R}} :e^{- i\hat \Phi(\xp, \xm)}:\, ,
}
where we introduced the full bosonic field
\eqn\Phixpxm{
\hat \Phi(\xp,\xm) =  \hat \Php(\xp)  + \hat \Phm(\xm) .
}
Since we performed a subtraction, the overall coefficient on the
r.h.s. of \Bilin\ was fixed by hand. This is done by comparing the
quasiclassical expressions obtained below through bosonization with
those that follow from the quasiclassical wave functions.

Using \Bilin, one can easily write the bosonization formula for any
observable of the form \operUU. In particular, for the trace of the
identity operator \trU, which gives the number of particles, one
obtains
\eqn\numbop{
N =  {1\over (2\pi)^{3/2}R}\int {d\xp d\xm
\over (\xp\xm)^{{R+1\over 2 R}  }}\, 
e^{i\xp\xm} \<\mu|:e^{- i \hat\Phi(\xp,\xm)}:| \mu\>.
}

\subsec{Quasiclassical limit}
 
\noindent
The quasiclassical asymptotics of the fermionic wave functions is
determined, through the representation \Boso, by the quasiclassical
expansion of the exponentials  of the bosonic fields $\hat \Phpm(\xpm)$.
Here we will restrict ourselves  to  the first two orders: 
\eqn\fercor{
\Pspm^{-\mu}(\xpm)= \frac{1}{{\sqrt{2\pi\, \xpm}}}\, 
 e^{ \mp i \Phi_{_{\pm}}(\xpm )   } \ 
e^{- {1\over 2} \langle : \hat\Phpm(\xpm ) \hat\Phpm(\xpm ) 
:\rangle_c}
,
}
where
\eqn\expvF{
\Phpm(\xpm )\equiv
\langle \hat\Phpm(\xpm )  \rangle 
}
is the vacuum expectation value of the field $\hat\Phpm(\xpm ) $
and $\langle \cdots \rangle_c$ denotes the connected correlator.
In the leading order the phase of the fermion wave functions 
are thus given by the expectation values of the  left and right bosonic fields:
\eqn\onept{
\vp_{_{\pm}}(\xpm;-\mu)+ \mu \log \xpm  \approx  \Phi_{_{\pm}}(\xpm) .
}
In particular, the quasiclassical phase of the fermion scattering is
equal to the expectation value of the zero mode $\hat\phi$
\eqn\phihatphi{
\phi = \langle \hat\phi\rangle.
}

In order to find the subleading factor in the exponential, we need the
two-point correlation functions. The connected
two-point correlators are generating functions for the derivatives
$\p_\mu \p_{n} \CF$ and $\p_n \p_m\CF$, which were calculated in the
case of sine-Liouville deformation in \refs{\AK} from Toda hierarchy.
The answer is actually true for any deformation and can be expressed 
in terms of the functions $\t(\xpm)$ obtained by inverting the
parametric representation \xpmo\ \KostovWV.\foot{In the case of a
compact Toda hierarchy, when the fermion eigenfunctions are entire
functions, these relations have been obtained in
\refs{\WiegmannXF,\TeodorescuYA}.} Note that the functions $\t(\xpm)$ 
are given by the derivative in $\mu$ of the phase \onept\ of the
fermion wave functions
\eqn\omtau{
\tau(\xpm) = \pm\p_\mu \Phpm(\xpm).
}
The explicit expressions of the generating functions are
\eqn\corrrr{\eqalign{
\hat \Dpm(\xpm)\, \p_{\mu}\CF\ \ \ \  &=-\frac{1}{2R}\,\p_{\mu}^2\CF
-\log\(\xpm \,e^{\mp \tau(\xpm)}\), \cr
\hat \Dpm(\xpm) \hat \Dpm (\ypm)\, \CF &=\ \frac{1}{2R^2}\,\p_{\mu}^2\CF+
\log\({\xpm^{1\over R}-\ypm^{1\over R}}\)
- \log\( {  e^{\pm {\tau(\xpm)\over R}}
-e^{\pm {\tau(\ypm)\over R}}}
\), \cr
\hat \Dp(\xp) \hat \Dm (\ym)\, \CF & =\   
\log\(1- e^{{\tau(\ym)\over R}-{\tau(\xp)\over R}}\).
}}
From the first two identities \corrrr\ one finds the connected correlation 
function for the left and right chiral fields:
\eqn\normordtp{\eqalign{
\<: \hat\Phpm(\xpm ) \hat\Phpm(\ypm ): \>_c=
\log\(\pm { \xpm^{1/R}-\ypm^{1/R}
\over 2(\xpm\ypm)^{{1\over 2R}}\sinh \frac{\t(\xpm) -\t(\ypm)}{2R}}\)
+\frac{1}{4R} \p_\mu\phi .
}}
Substituting this in \fercor, one obtains 
the quasiclassical form of the fermion wave functions:
\eqn\quasboswf{
\Pspm^{-\mu}(\xpm)=
{e^{ \mp i \Phpm(\xpm )-{1\over 8R}\p_\mu\phi}
\over \sqrt{\pm 2\pi \p_\tau \xpm}}.
}
Up to the last term in the exponent, the expression \quasboswf\
reproduces the previously obtained result \quaswf\ together with \phiint.
 
To explain the appearance of the last term, let us consider the zero
modes $\phi_\pm$.  To all orders, they are given by
\eqn\phases{
e^{\mp i \phi_\pm} = \langle e^{\mp i \hat \phi/2} \rangle
= {\CZ\(\mu\mp {i\over 2R}\)\over \CZ(\mu)},
}
so that for the total zero mode we reproduce the expression \FRENOI:
\eqn\tatlphs{
e^{i\phi}=e^{i (\phi_++\phi_-)}=
{\CZ\(\mu+{i\over 2R}\)\over \CZ\(\mu- {i\over 2R}\)}.
}
On the other hand, the quasiclassical calculation gives
\eqn\corphase{
e^ {\mp i \phi_\pm} \approx   
e^{\mp i \< \hat\phi/2\>- {1\over 8} \< \hat\phi\hat\phi \>_c}
= e^{ \mp i \phi /2 - {1\over 8R}\p_\mu \phi},
}
which is, of course, agrees with the expansion of \phases.  In this
way the last term appears as a correction to the zero modes of the
wave functions of chiral fermions. This term reflects the freedom in
the choice of the relative normalization of the left and right
fermions and is cancelled in the product $\overline{\Psm^{-\mu}}\Psp^{-\mu}$.

Now let us turn to the fermion bilinear \Bilin. In the leading order,
its expectation value is given by the exponential of  the expectation 
value of the full bosonic field,
\eqn\expPpm{
\langle \hat\Phi (\xp, \xm)\rangle = \Php(\xp)+ \Phm(\xm).
}
 To find the subleading contribution, one
should again evaluate the two-point correlation function of
$\hat\Phi(\xp,\xm)$. From \corrrr\ one finds
\eqn\twN{\eqalign{ 
\<  :\hat\Phi(\xp, \xm) \hat\Phi(\yp, \ym): \>_c= &
\log{ \sinh \frac{\t(\xm) -\t(\yp)}{2R}  \, \sinh \frac{\t(\ym)-\t(\xp) }{2R}
\over \sinh \frac{\t(\xp) -\t(\yp)}{2R} \, \sinh \frac{\t(\xm)-\t( \ym) }{2R} }
\cr
+&  \log (\xp^{-1/R} -\yp^{-1/R})+\log (\xm^{-1/R}-\ym^{-1/R}),
}}
which yields for the expectation value of the fermion bilinear
\eqn\ferbilcor{\eqalign{
\< \mu|\hat\Psm^{\dagger}(\xm) \hat\Psp(\xp)|\mu  \>\ & =\ 
\frac{1}{2\pi R}\, ( \xp\xm)^{-{R+1\over 2R} }\,
 e^{- i  \Phi(\xp, \xm ) 
-{1\over 2}\<:\hat\Phi(\xp, \xm )\hat\Phi(\xp, \xm ):\>_c }
\cr 
& =  
{e^{- i \Phi(\xp, \xm) }\over 2\pi \sqrt{\p_\t \xp \p_\t \xm}}\
{1  
\over 2R\, \sinh \frac{ \t(\xm) -\t(\xp) }{2R} }.
}}

The last two terms on the r.h.s. of \twN\ appear due to the normal 
ordering. Without normal ordering the correlation function of 
the gaussian field would be given only by the first term. Taking this into account,
the formula \ferbilcor\ generalizes straightforwardly
to the case of several fermion bilinears:
\eqn\manyFZZ{
\langle \mu| \prod_{j=1}^n \hat\Psm^{\dagger}(\xm^j)\hat\Psp(\xp^j) |\mu \rangle
=\prod_{k=1}^n  { e^{-i  \Phi(\xp^k, \xm^k) }
\over 2\pi\sqrt{\p_\t \xp ^k\p_\t \xm^k}
}\under{\det}{i,j}\( {1\over 2 R\,\sinh  \frac{ \t(\xm^i)  -\t(\xp^j) }{2R} }\).
}
We expect that this formula becomes exact if one replaces the function
$\o_\pm = e^{\t(\xpm)}$ by the shift operator $e^{i\p_\mu}$ of the
Toda hierarchy.\foot{In the case of the two-matrix model, where the
integrable structure is that of KP hierarchy, such an exact formula
was derived in \MaldacenaSN.}

Note that the second factor in the denominator of \ferbilcor\ diverges
on the surface of the Fermi sea where $\tau(\xp)=\tau(\xm)$. This
divergence appears due to breakdown of the quasiclassical
approximation of the bosonization formulae near the Fermi sea.  This
divergent factor can be canceled by application of a difference
operator in $\mu$. Indeed, in the given approximation such an
operator acts only on the exponent. Therefore, applying the relation
\omtau, one concludes that
\eqn\difcc{
2R\sin(\frac{1}{2R} \p_\mu) 
\< \mu|\hat\Psm^{\dagger}(\xm) \hat\Psp(\xp)|\mu  \>=
-{e^{- i \Phi(\xp, \xm) }\over 2\pi\sqrt{-\p_\t \xp \p_\t \xm}}.
}
As a result, the r.h.s. becomes the quasiclassical
expression for the product of the wave functions
$\overline{\Psi^{-\mu}_-(\xm)}\Psi^{-\mu}_+(\xp)$. This allows to
write a difference equation for the expectation value of any operator
$\CO_f$ of the form \operf\
\eqn\FRBZ{ 
2R\sin(\frac{1}{2R} \p_\mu) \<  \mu | \hat\CO |\mu\>= 
-\frac{1}{ (2\pi)^{3/2} }\int \frac{d\xp d\xm}{   \sqrt{-\p_\t \xp \p_\t \xm} }     \,  
f(\xp,\xm) \, {e^{ i\xp\xm- i    \Phi(\xp, \xm) }}.
}

The non-perturbative effects to the free energy follow from this result if one  
specify $\CO$ to be the identity operator $(f=1)$. 
Then the relation \sintr\ implies
\eqn\normbos{
\frac{1}{\sqrt{2\pi}}\int d\xp d\xm {e^{ i\xp\xm- i  \Phi(\xp, \xm) }
\over \sqrt{-\p_\t \xp \p_\t \xm}  }
= \log\L.
}
The evaluation of this  integral by the saddle point method gives 
the non-perturbative corrections to the zero mode $\phi$ and, through
the relation \FRENOI, to the free energy itself, as it was done in
section 4.3.

\newsec{Discussion}

In this paper we studied non-perturbative corrections to the partition 
function of the compactified $c=1$ string theory deformed by a generic tachyon source.
The flows between different backgrounds are described by a non-compact
realization of Toda hierarchy.
The space of all such backgrounds is parametrized by a set of Toda 
`times' $\{t_{\pm k}\}$, which in the world sheet CFT  
have the meaning of coupling constants for the allowed marginal deformations.
Each such time-dependent background is described by a complex curve, 
which gives the solution of Toda hierarchy in the dispersionless, 
or quasiclassical, limit.

We studied the dual realization of the string theory as the singlet 
sector of the matrix quantum mechanics, described by free fermions in
upside-down oscillator potential, using chiral canonical coordinates
$\xpm \sim x\pm p$. In the fermionic system the complex curve appears as
the complexified classical trajectory at the Fermi level. This curve 
has an infinite (for $R$ irrational) number of double points, 
obtained as solutions of a transcendental equation.
The leading non-perturbative corrections are associated with 
these double points, in the same way as in the minimal string theories.
The exponents for the leading NPC are given by integral of a 
holomorphic differential along the compact cycles associated with these double points.

Our main new result is the expression for the subleading correction given 
in \Akk\ for a general tachyon potential and in \Afsolff\ 
for the particular case of sine-Liouville theory.
We derived these results by quasiclassical analysis of the
fermionic wave functions as well as by solving the linearized Toda
equation in the simplest case of sine-Liouville deformation.
Then we showed that these results have a natural interpretation in 
terms of a bosonic collective field, whose oscillator modes are  
in correspondence with the Toda couplings $t_{k}$. The fermion 
bilinears then can be represented as vertex operators of this bosonic field, 
eq. \Bilin. The subleading corrections are proportional to the exponent 
of the two-point bosonic correlation function, eq. \ferbilcor.

It is natural to expect that the NPC have their origin in the $D$-branes 
of the string theory. In order to have such an interpretation, 
they should be expressed in terms of world sheets with
local boundary conditions. Such an interpretation exists for the
eigenvalue $x$ of the random matrix variable, which is the boundary
cosmological constant for the Liouville field.

On the other hand, our results were derived in terms of the chiral phase-space coordinates, 
$\xp$ and $\xm$, which do not have direct interpretation in terms of local conformal 
boundary conditions in the world-sheet theory.  
Therefore, to find the meaning of the non-perturbative corrections as the effects 
due to D-branes, we should first translate the results obtained in the $(\xp, \xm)$ 
representation back to the $(x,p)$ representation.

Assuming that  $t_k=t_{-k}$ and that there is a finite number 
of non-vanishing couplings, the parametric form of the curve  
in the $(x,p)$ space is obtained directly from \xpmo:
\eqn\pxo{\eqalign{
x(\t) &= \sqrt{2}\, e^{- {1\over 2R} \chi} \Bigl[ \cosh \t
+ \sum_{k=1}^{k_{\max}}  a_k  \cosh \( (1-\frac{k}{R}) \t\) \Bigr], \cr
p(\t) &= \sqrt{2}\, e^{- {1\over 2R} \chi} \Bigl[ \sinh \t
+\sum_{k=1}^{k_{\max}}  a_k  \sinh\( (1-\frac{k}{R}) \t\) \Bigr].
}}
The  FZZT curve is the one for the  resolvent  $y=w(x)$.
The latter is related to the  momentum $p$  by 
\eqn\pofx{
p(x) = {w(e^{ i\pi} x) - w(e^{-i\pi } x)\over 2 i}.
}
This equation is easy to solve only  in the case of stationary background, 
where the two sides of the cut of $w(x)$ are the images of  the 
straight lines $\Im \t = \pm \pi$, and whose solution is given by
\freepxW.
In the case of a time-dependent background the two sides of the 
cut of $w(x)$ are parametrized by the lines
$\t_\pm (t) =t \pm i \pi_1(t), \ t>0$, where $\pi_1( t)\to\pi$
only asymptotically at $ t\to \pm \infty$.
Therefore  eq. \pofx\ can be written more explicitly as
\eqn\pofxt{
p(\t) = {w( \t + i\pi_1(\t)) - w( \t - i\pi_1(\t)) \over 2i}, 
} 
where the function $\pi_1( t)$
is obtained as the solution of the transcendental equation
\eqn\sheetse{
\sin\pi_1 + \sum_{k=1}^{k_{\max} } a_k { \sinh \((1-\frac{k}{R}) t\)
\over \sinh  t}\, \sin \((1-\frac{k}{R}) \pi_1\)=0,  
\quad t \in {\Bbb R}.
}
As a consequence, the parameter $\t$ does not uniformize
the curve $y=w(x)$. In general, $w(\t)$ will have infinitely
many branch points.

Considered at $ t = 0$, eq. \sheetse\ coincides with the condition that  
$\p_\t x=0$  for purely imaginary $\t$.  
Therefore  the point  $\t = \pm i\pi_1(0)$ is a branch point  for $p(x)$.
It is easy to see that all purely imaginary  solutions 
of \sheetse\ correspond to branch points of $p(x)$.
The solutions of \sheetse\ can be classified by their asymptotics 
at infinity: $\pi_n(t)\to n \pi$.

The equation \sheetse\ for $\pi _n(\t)$ is not compatible with 
the transcendental equation \thetak\ for the parameters $\theta_n$ 
of the double points, except for the stationary background.  
Therefore, for time-dependent backgrounds the double points never occur 
at the branch points of the resolvent or along its cuts.  
This fact can be geometrically understood by considering the classical 
trajectory  $x(i\th)$ in imaginary time direction.
The branch points are the turning points of the classical trajectory,
while the double points are associated with its intersection points.  
Therefore, turning on a time-dependent tachyon background 
resolve the infinite degeneracy of the double points of the 
complex curve for the $c=1$ string theory.
       
Now let us return to the interpretation of the non-perturbative corrections 
in terms of D-branes.
For the leading non-perturbative effects, which are given by the 
closed contour integrals $S_n$ \contgk, this was done in
\refs{\SAcurve,\KazakovDU }. 
Using the relation (D.10) of the light-cone coordinates
to the spectral density and \pofx, one obtains
\eqn\Skres{
S_n= -\frac{1}{2}  \oint_{\g_n} dx \(w(e^{i \pi } x)-w(e^{- i \pi } x)\).
}
The resolvent $w(x)$ is equal to the derivative  
of the FZZT disk partition function with respect
to the boundary cosmological constant $x$:
\eqn\FZZres{
w(x)= \p_x \Pfzz(x).
}
Taking into account the symmetry  $w(e^{i\pi}x(\tau))=w(e^{-i\pi}x(-\tau))$
of the resolvent, this allows to identify
\eqn\SkFZZ{
-S_n =\Pfzz(e^{-i\pi}x(i\theta_n))-\Pfzz(e^{i\pi}x(i\theta_n)). 
}

In \SAcurve\ it was also checked in the first order in the deformation
coupling $\l$
that the leading correction is given by the ZZ disk partition function 
\eqn\ZZlead{
-S_n =\Phi_{_{ZZ}}(n,1)
}
so that the $n$-th double point of the complex curve is associated to   
the $(n,1)$ ZZ brane. 
Assuming that this equation remains to be true in the full deformed theory,
from \SkFZZ\ and \ZZlead\ we conclude that the ZZ partition
function is equal to the difference of two FZZT partition functions
calculated
from two sides of the double point. 
This is a generalization to the case of a general tachyon  
deformation of a similar identity in the non-deformed theory $(\l=0)$.
\refs{\MartinecKA,\Hos,\TeschnerRim,\Pons}.

Note that our analysis also reveals the distinguished role of the
$(1,1)$ ZZ brane.  Up to now it was the only one which played a role
in the interpretation of matrix models as theories of open strings.
In our context it is associated with the first double point of the
complex curve.  One can show that this is the only double point which
is situated on the physical sheet of the Riemann surface.  It is
natural to expect that this fact corresponds to the observation in CFT
that the $(n,1)$ ZZ branes with $n>1$ possess some pathological
properties \ZamolodchikovAH. However, if we are calculating the
non-perturbative corrections to observables that explore lower sheets
of the Riemann surface, than the other double points may become
relevant.

The world-sheet interpretation of the subleading order of NPC seems to
be a more complicated problem. It is expected to be related to the
annulus amplitude on ZZ brane. However, the expression
given for this amplitude in \KutasovFG\ diverges in this case.  One
may attempt to relate this divergence to the factor
$\(\log\Lambda\)^{-1}$ in \coefCCC. But in this way we can not explain
the universal non-trivial functional dependence on the couplings.
On the other hand, our results  are naturally expressed in
terms of  the two-point  amplitude of a gaussian field whose left 
and right components are associated with the incoming and outgoing tachyons.
It is not however obvious how to reformulate them in terms of the 
annulus open string amplitude on FZZT branes,  
which is defined in terms of the eigenvalue $x$.

\smallskip\smallskip\smallskip
\bigskip
\noindent
{\bf  Acknowledgments}
\smallskip

\noindent
The authors would like to thank V. Kazakov, M. Kleban, R. Rabadan, 
O. Ruchayskiy, N. Seiberg, D. Shih, M. Taylor and especially 
S. Vandoren and P. Wiegmann for many valuable discussions.
I.K. thanks the Institute for Advanced Study, Princeton, for their 
kind hospitality during part of this work.
This research is supported in part by the
European network EUCLID, HPRN-CT-2002-00325.

\appendix{A}{Relations between the density of states, the scattering phase and the free energy}

\noindent
Knowing the zero mode $\phi(E)$ of the fermion wave functions,
one can reconstruct the density of states by  introducing
a completely reflecting cut-off wall at distance $\Lambda\gg\mu$\
\refs{\AKKNMM, \KostovTK}. The wall introduces a boundary condition
at $\xp=\xm=\sqrt{\Lambda}$
\eqn\bcctff{
[\hat S \Psi](\ctf) =\Psi(\ctf).
}
Thus, one identifies the scattered state with the initial one at the wall.
Applying this condition to the deformed wave functions \asswave,
one obtains that it is satisfied for a discrete set of energies $E_n\ (n\in \Z )$
defined by
\eqn\kvconbis{
\phi(E_n) - E_n\log \Lambda+V(\Lambda)+2\pi n=0,
\qquad V(\Lambda)=\sum\limits_{k}(t_k+t_{-k})\Lambda^{k/2R}.
}
From \kvconbis\ one can find the density of the energy levels in the confined system
\eqn\DEN{
\rho(E)= {\log \Lambda\over 2 \pi}-
{1\over 2\pi } {d\phi(E)  \over d E} .
}

Now the free energy $\CF(\mu, R)= \log\CZ(\mu, R)$
of the fermionic system compactified at distance $\b = 2\pi R$
can be calculated using \FRENO\ with the density \DEN.
Dropping out the $\Lambda$-dependent non-universal contribution and
integrating by parts, one obtains
\eqn\FRENOo{
\CF(\mu) =-{1\over 2\pi}\int d\phi(E)
\log \left( 1+e^{-\beta(\mu+ E)}\right)=
- R \int_{-\infty}^{\infty}dE {\phi(E) \over 1+e^{\beta(\mu+ E)}} .
}
We close the contour of integration in the upper half plane
and take the integral as a sum of residues.
This gives for the free energy
\eqn\FRENOoo{
\CF(\mu) =i \sum _{r=n+\hf > 0}
\phi\left( i r/R-\mu\right)
}
from which \FRENOI\ follows.

\appendix{B}{Solution of \eqep}

\noindent
In this appendix we solve the equation \eqep\ using the ansatz \anz.
Substituting \anz\ into \eqep, keeping only the leading terms in
the $\xi\to 0$ limit (which come from the terms with all or all except one
derivatives acting on the exponent in \anz), and taking into account that
$g(y)$ satisfies the equation
\eqn\eqg{
\sqrt{\alpha}\,e^{{X(y)\over 2R^2} } (1- y\p_y) g(y)
=\sin\[\oR\p_y g(y)\],
}
one finds the following first order differential
equation on $A(\xi,y)$:
\eqn\eqA{\eqalign{
&\alpha\,e^{{X\over R^2}} \left[ 2\((1- y\p_y) g\)
\( y\p_y+\xi\p_{\xi}\) \log A- (1-y\p_y)^2 g \]
\cr
&=-\oR\sin \frac{2g'}{R}\, \p_y \log A-\frac{g''}{R^2}\cos\frac{2g'}{R} .
}}
Taking into account the form of $A(\xi,y)$, one obtains two equations
for $a(y)$ and $b(y)$
\eqn\eqab{\eqalign{
& b'\left[ 2\alpha\,e^{{X\over R^2}} (1- y\p_y) g + \oR\sin\frac{2g'}{R}\right] =0, \cr
& \alpha \, e^{{X\over R^2}}\[2b\( (1- y\p_y) g \) \(1+ y\p_y \log a\)-
(1-y\p_y)^2 g\] = 
-\frac{b}{R}\sin\frac{2g'}{R}\, \p_y \log a-\frac{g''}{R^2}\cos \frac{2g'}{R} .
}}
The only solution of the first equation is
\eqn\bsol{
b=const.
}
Then the solution of the second equation is given by the following
integral
\eqn\asol{
\log a(y)=\int dy\,  {
b^{-1}\[\alpha \, e^{{X\over R^2}}(1-y\p_y)^2 g-{g''\over R^2}\cos {2g'\over R}\]
-2\alpha \, e^{{X\over R^2}}(1-y\p_y)g
\over 2\alpha\, e^{{X\over R^2}} y(1-y\p_y)g+\oR\sin{2g'\over R} }.
}
Using \eqg, the integral can be rewritten as
\eqn\asoli{
\log a(y)=-{1\over 2b}
\log \left|\alpha\, e^{{X\over R^2}}y(1-y\p_y)g+\frac{1}{2R} \sin\frac{2g'}{R}\right|+
\int dy\,  { \sqrt{\alpha} e^{{X\over 2R^2}}
\( {1\over 2b}(2+{y\over R^2}\p_y X)-1\)
\over \oR\cos {g'\over R} -2\sqrt{\alpha} \sinh \( {2R-1\over 2R^2}X\) }.
}
Then we change the integration variable to $X$ using the equation \wf.
The second term in \asoli\ becomes
\eqn\asolii{
{\sqrt{\alpha}}\int dX\,  {
- {2R-1\over bR} \cosh \( {2R-1\over 2R^2}X\)
+\[ (\oR-1) e^{{2R-1\over 2R^2}X} -e^{-{2R-1\over 2R^2}X}\]
\over \oR\cos {g'\over R} -2\sqrt{\alpha} \sinh \( {2R-1\over 2R^2}X\)  }.
}
Changing the variables to $\theta =g'$ by means of \zzz, after simple algebra
one arrives at the following integral
\eqn\asolphi{
\eqalign{
  - &\int d\theta \,
 { {1\over b} \[ \sin^2 \( {1-R\over R}\,\theta \)+\(\oR-1\)\sin^2{\theta}\]
+ {2R\over 2R-1}\[\(\oR-1\)^2 \sin^2{\theta }-\sin^2\( {1-R\over R}\,\theta \)\]
\over
\sin\theta  \sin{\theta \over R}\sin \( {1-R\over R}\,\theta \)  }
  \cr
=& {1\over b}\log\[{C\, \sin{\theta \over R} \over
\sin{\theta}\sin \( {1-R\over R}\,\theta \)}\]
+{2R\over 2R-1}\log\[{\sin{\theta }\over \(\sin{\theta \over R} \)^{2-\oR}
\(\sin \( {1-R\over R}\,\theta \)\)^{\oR-1} }\],
}}
where $C$ is an integration constant.
Combining this result with the first term in \asoli\ rewritten in terms
of $\theta $, one obtains
\eqn\aresult{\eqalign{
&
\log a(y)=\frac{X}{R}-2\log\[ { \sin{\theta \over R} \over\sin \( \frac{1-R}{R}\,\theta \)}\]
\cr
&-\frac{1}{2b}
\log\[\(\oR-1\)\cot\( \frac{1-R}{R}\,\theta \)-\cot {\theta } \]
-\frac{1}{b} \log\[C^{-1}\sin{\theta }\sin \( \frac{1-R}{R}\,\theta \)\].
}}
As a result, the pre-exponential factor reads
\eqn\Asol{
A(\xi,y)=C\[\xi\,
{e^{{X\over R}}\sin^2 \( {1-R\over R}\,\theta \)\over\sin^2{\theta \over R} }\]^b
\left\{ \sin^2{\theta}\sin^2 \( \frac{1-R}{R}\,\theta \)
\(\(\oR-1\)\cot\( \frac{1-R}{R}\,\theta \)-\cot {\theta}\)\right\}^{-1/2}.
}

\appendix{C}{The $c=0$ critical limit for the subleading contribution}

\noindent
It is well known \HSU\ that sine-Liouville theory
exhibits a critical behavior when sine-Liouville coupling
becomes sufficiently large. Physically, at the critical point
the fluctuations of the matter field get frozen in minima of sine
potential. As a result, the system approaches the $c=0$ CFT describing
pure two-dimensional gravity.

In \refs{\KAK,\SAmn} it was shown that
the leading non-perturbative correction to the partition function corresponding
to $n=1$ follows the same pattern, whereas other corrections with $n>1$
disappear in the $c=0$ critical limit.
Thus, we expect that $A_1(y)$ must exhibit a singularity as $y\to y_c$ with
\eqn\yyccrr{
y_c=-(2R-1) R^{-{R\over 2R-1}}(1-R)^{1-R\over 2R-1}
}
and, furthermore, the behavior of $A_1$ near this singularity should reproduce
the non-perturbative effects of the $c=0$ theory
\eqn\Acrit{
A_1\sim  g_{{\rm str},c=0}^{1/2}
\sim (y-y_c)^{-5/8}.
}

Indeed, the singularity at $y=y_c$ corresponds to a critical point of \wf,
near which the relation between $y$ and $X$ degenerates:
\eqn\yXrel{
{y_c-y \over y_c}\simeq {1-R\over 2R}(X-X_c)^2+O\((X-X_c)^3\).
}
At the critical point $\theta _1\to 0$ and
the first two terms in the expansion of $\theta _1$ around the singularity are
\eqn\crphi{
\theta _1(y)={\sqrt{3}}(X_c-X)^{1/2}-{\sqrt{3}(2R^2-2R+1) \over 20 R^3}(X_c-X)^{3/2}
+O\((X_c-X)^{5/2}\)~.
}
Due to this result, the leading term in \Asolf\ is
\eqn\Acr{
A_1(y)\approx C\, {R^{3R-2\over 2R-1} \((1-R)\lambda\)^{-{R\over 2R-1}}
\over 3^{3/4}\sqrt{2R-1}}\,(X_c-X)^{-5/4}.
}
Taking into account \yXrel, one finds the law \Acrit\  corresponding
to the $c=0$ theory.

\appendix{D}{Quasiclassical wave function in the $x$-representation}

\noindent
In this appendix we study the quasiclassical wave function of the
deformed fermionic system in the $x$-representation where $x$ is the
usual fermion coordinate related to eigenvalues of MQM. The simplest
way to get this function is to use the result \quaswf\ together with
\phiint\ for the same wave function in the light-cone representation.
Then it is enough to apply a unitary operator relating the two
representations.

This unitary operator can be represented as an integral operator with
kernel defined as a solution of the following equations
\eqn\kereqpm{
{1 \over\sqrt{2}}\(x\mp i\p_x \) |x \rangle \langle \xpm| =\xpm |x \rangle \langle \xpm|,
}
\eqn\kereqx{
{1 \over \sqrt{2}}\(\xpm\pm i\p_{\xpm} \)\(|x \rangle \langle \xpm|\)^* =
 x \(|x \rangle \langle \xpm|\)^*.
}
It is easy to check that it is given by
\eqn\kernn{
|x \rangle \langle \xpm|={1\over 2^{1/4}\sqrt{\pi}}\,
e^{\mp{i\over 2}\( x^2-2\sqrt{2}x\xpm+\xpm^2\)}.
}
Then one obtains (the same result will be for the ``$\xm$" representation)
\eqn\wavefunx{\eqalign{
\Psi^{_{E}}(x)&=
\int_{0}^{\infty} d\xp \, |x \rangle \langle \xp|\Psi^{_{E}}_{+}(\xp)
\cr
&= \int_{0}^{\infty} d\xp \,{\,e^{- i \phi_{+}}
\over 2^{3/4} \pi\sqrt{\p_{\tau}\xp}} \,
e^{-{i\over 2}\( x^2-2\sqrt{2}x\xp+\xp^2
+2\int_{\infty}^{\xp}\txm(\pxp)d\pxp\)}.
}}
The integral can be calculated by the saddle point method.
The saddle point equation reads
\eqn\sadpeqx{
{\xp +\txm(\xp)\over \sqrt{2}}=x,\quad {\xp -\txm(\xp)\over \sqrt{2}}=p
}
where $\txm(\xp)$ is the function defined by \xpmo. The fluctuations of the phase
around this point are $-\(1+{d\xm \over d\xp}\){(\delta\xpm)^2\over 2}$. Altogether
these results give
\eqn\wfxx{
\Psi^{_{E}}(x)={1\over\sqrt{2\pi|\p_{\tau}x}|}\,
\exp\({{i}\int^x p(x')\,dx'}\),
}
where the low limit of integration in the phase can be chosen to cancel the zero mode
$\phi_+$.

Finally, let us show how the relation, which is evident in the classical limit,
between the density $\vr(x)$ of eigenvalues and the functions $\xpm=\txpm(\xmp)$
follows from the quasiclassical wave function \wfxx.
The density in the compactified theory can be written in terms of the fermion
wave functions as follows
\eqn\densxx{
\vr(x)=2\int\limits_{-\infty}^\infty \!\! dE\,
{\bar \Psi^{_E} (x)\Psi^{_E} (x)\over e^{-\b(\mu+E)}+1},
}
where the coefficient 2 comes from the fact that there are left and
right moving fermions.  Substituting the quasiclassical limit \wfxx\
for the wave functions and replacing the thermal factor by
$\th(-\mu-E)$, one finds
\eqn\derrho{
\p_\mu \vr (x)= -{1\over  \pi |\p_\t  x|} .
}
From here, using the identities   
\eqn\propder{
\frac{ \p\xmp }{ \p E}\Big|_{_{\xpm}}=\mp \frac{1 }{ \p_{\tau}\xpm}, 
\qquad \p_{\mu}\big|_{_{x}}=\p_{\mu}\big|_{_{\xp}}-
\frac{1}{\sqrt{2}\p_{\tau}x}\p_{\xp}\big|_{_{\mu}} ,
}
one gets the well known result for the quasiclassical spectral density
\eqn\densx{
\vr(x) = \frac{1}{\pi\sqrt{2}} \( \xp-\xm\), \qquad x =  \frac{1}{\sqrt{2}} \(\xp+\xm\).
}

\listrefs

\bye